\documentclass[conference]{IEEEtran}
\usepackage{hyperref}
\usepackage{mathpazo}
\usepackage[utf8]{inputenc}
\usepackage[T1]{fontenc}
\usepackage{tabularx}
\usepackage{lscape}
\usepackage{placeins}
\usepackage{longtable}

\usepackage{amsmath}
\usepackage{graphicx}

\usepackage{comment}
\usepackage{tabularx}
\usepackage{ltablex}
\usepackage{array}

\usepackage{float} 
\excludecomment{comment} 

\begin{document}

\title{Wi-Fi Channel Saturation as a Mechanism to Improve Passive Capture of Bluetooth Through Channel Usage Restriction}

\author{Ian Lowe, William J Buchanan, Richard Macfarlane, Owen Lo\\The Cyber Academy, Edinburgh Napier University, Edinburgh}

\maketitle

\begin{abstract}
Bluetooth is a short-range wireless technology that provides audio and data links between personal smartphones and playback devices, such as speakers, headsets and car entertainment systems. Since its introduction in 2001, security researchers have suggested that the protocol is weak, and prone to a variety of attacks against its authentication, link management and encryption schemes. Key researchers in the field have suggested that reliable passive sniffing of Bluetooth traffic would enable the practical application of a range of currently hypothesised attacks. Restricting Bluetooth’s frequency hopping behaviour by manipulation of the available channels, in order to make brute force attacks more effective has been a frequently proposed avenue of future research from the literature. This paper has evaluated the proposed approach in a series of experiments using the software defined radio tools and custom hardware developed by the Ubertooth project. The work concludes that the mechanism suggested by previous researchers may not deliver the proposed improvements, but describes an as-yet undocumented interaction between Bluetooth and Wi-Fi technologies which may provide a Denial of Service attack mechanism.
\end{abstract}
 
\section{Introduction}
Bluetooth describes a communications environment consisting of radio hardware, protocol stack, and service implementation in a similar usage as the term \emph{Web} describing the entire Internet ecosystem.

Originally developed as an internal project by Dr Jaap Haartsen of Ericsson Mobile \cite{Haartsen2000}, Bluetooth was offered to industry through the Bluetooth Special Interest Group (SIG) in 1998.

The SIG published the Bluetooth specification in 1999, and within two years theoretical weaknesses had been described by researchers. Jacobsson and Wetzell \cite{Jacobsson2001} suggested that a potential attack against the pairing mechanism might allow link keys to be recovered, and in a second weakness, poor choice of keys reduced the effective strength of the cipher. Despite these potential weaknesses, the standard was widely adopted by mobile phone manufacturers, as a means of connecting to audio headsets and emerging data devices. 

The near ubiquity of Bluetooth support for mobile phone applications prompted automobile manufacturers to implement support for the standard \cite{Heffernan2001}. Current implementations of Bluetooth in an automotive environment provide deep integration between in-car information and entertainment systems, vehicle systems and the driver’s smartphone \cite{checkoway2011comprehensive}.

This integration provides considerable utility; however,recent work has highlighted a variety of possible attacks, and Bluetooth based attacks feature heavily in this research \cite{cheah2017towards}. Because it uses a radio medium with authentication and encryption schemes with identified weaknesses, Bluetooth networks are perceived to be weak. 

Three broad classes of attack have been described:

\begin{enumerate}
\item Attacks against Bluetooth services and applications, making use of weaknesses in the authentication and authorisation processes;
\item Attacks using information transmitted by the device for unauthorised tracking of the user’s location or behaviour; and
\item Attacks which seek to intercept traffic to gain access to voice calls and other, private information.
\end{enumerate}

This paper focuses on the third class of proposed attacks. Taking this further, the focus is on passive sniffing – eavesdropping traffic without connecting to the devices in question. Passive sniffing in this way has been frequently hypothesised \cite{Shaked2005} and researchers have described potential mechanisms \cite{Spill2007}, \cite{Huang2014}. The remainder of this paper details the approaches that have been taken, the progress made towards the goal of passive sniffing, and seeks to experimentally evaluate the extent to which reducing available bandwidth through active manipulation of Adaptive Frequency Hopping (AFH), can be used to reduce the time required for brute force attacks, and therefore support passive sniffing of Bluetooth.

\begin{comment}

\subsection{Layout of this Document}
Following this introduction, Section 2 of the paper sets out key contributions from the published literature. As an analysis of the implementation of a technical standard, relevant information from the Bluetooth Standards in version 2.1, version 4.1 and version 5.0 is used to explain the operation of relevant features of the protocol. The various weaknesses identified in the implementation of these features, and subsequent development of models of attack is identified.

An incorrect assumption that Bluetooth researchers make – that the underlying Frequency Hopping behaviour of the system does not add significantly to the protocol’s security – is introduced, and key examples in the literature highlighted.

Having laid out the terminology and key concepts required, Section 3 outlines the experimental methodology adopted, explaining the tools used and scenarios tested. It discusses the reasons for the choice of given technologies, and the steps taken by the author to eliminate possible sources of external influence.

In Section 4, the results of experimentation are considered – evaluating the success and failure of given approaches, the limitations of the tools and techniques used, and the relevance of the findings in relation to those publications examined in Section 2.

Finally, conclusions are summarised in Section 5, which also includes proposals for future avenues of study in the area, building upon the work carried out in this paper.
\end{comment}

\section{Literature Review}

Dunning \cite{Dunning2010} provides a taxonomy and classification of Bluetooth attacks as a series of hierarchies of classification, threat level and party responsible for mitigation – vendor, or end user. Of the 45 attacks identified in the survey only three target the PHY, MAC or LLC Layers (Table \ref{tab:1}). 

\begin{table*}[t]
  \caption{Dunning's Classification scheme (adapted from \cite{Dunning2010})}
 \begin{tabular}{l | l | l}
Classification	& Tools &  Attacks \\
\hline 
Surveillance	& btaudit, sdptool, Bluescanner, BTScanner	& RedFang, BlueFish, Blueprinting, War-nibbling\\
Range Extension	& Vera-NG	& BlueSniping, bluetooone\\
Obfuscation	& Hciconfig, bdaddr	& Spooftooph\\
Fuzzer	& &	BluePass, Bluetooth Stack Smasher, BlueSmack, Tanya, BlueStab\\
Sniffing &	BlueSniff, HCIDump, Wireshark, Kismet &	FTS4BT, Merlin\\
Denial of Service	& &	Battery exhaustion, signal jamming, BlueSYN, Blueper, BlueJacking, vCardBlaster\\
Malware	& &	BlueBag, Caribe, CommWarrior\\
Direct Data Access &	BlueSnarf, BlueSnarf++ &	Bloover, BlueBug, Car Whisperer, HeloMoto, btpincrack\\
Man in the Middle &	Bthidproxy &	BT-SSP-Printer-MITM, BlueSpooof\\
\end{tabular}
  \label{tab:1}
\end{table*}

This pattern is repeated in the attacks against Bluetooth described by Haines \cite{Haines2010} – of the seven attacks he describes, only one is not included in Dunning’s review. This additional attack btCrack is a sniffer which is based on the HCIDump tool, and attempts to recover link keys from a captured data stream. Haines is unique among these researchers, observing that sniffing a suitable stream of packets in the first place is significantly harder than in Wi-fi.

\begin{comment}
The more recent work of Hassan \cite{Hassan2017} places most of these attacks in a formal hierarchy; they identify some thirty attacks on Bluetooth. 

Most of these are previously described by Dunning and Haines; however, they describe a small number of malware programs. In each case, these practical attacks target the higher level layers.  Most recently of all, \cite{cheah2017towards} has produced an updated version of Dunning’s taxonomy reflecting the current threat environment. Whilst Cheah et al do not describe individual attacks, the taxonomy which they have constructed reinforces that attacks on the higher protocol layers represent the bulk of those described.
\end{comment}

\subsection{PHY and MAC - Bluetooth as an RF System}
Bluetooth devices communicate using Radio Frequency signals in the 2.4GHz Industrial, Scientific and Medical (ISM) Band. This is an internationally agreed allocation of spectrum which is intended for devices which can be operated without a user licence. 
The RF and Baseband systems within Bluetooth devices are not typically implemented by a device manufacturer – this core functionality is implemented in proprietary chipsets or System on Chip (SOC) components from leading vendors such as Qualcomm, Texas Instruments, Microchip etc. \cite{Chokshi2010} and the firmware of these devices is not open sourced \cite{Huang2014}.  

\cite{Pelzl2006} provide a detailed explanation of the RF layer of the Bluetooth Classic environment. Frequency Shift Keying (FSK) is a modulation scheme that uses the change between two distinct frequencies within the allocated band to represent a digital 0 and 1. In its simplest form, Bluetooth BR, or Basic Rate, uses a modified form of this scheme - Gaussian Frequency Shift Keying (GFSK). 

\begin{comment}The modulated signal is passed through a Gaussian filter, which has the effect of smoothing the transition from one frequency to the other. This reduces the amount of cross-band interference and is another conscious design adaptation to ensure that Bluetooth devices can operate effectively in noisy, congested radio environments.
\end{comment} 

This choice of modulation scheme has the unintended side effect of adding a further degree of complexity to the process of sniffing traffic \cite{Ossmann2009}. As the transitions between encoded digits are less precise, a potential attacker attempting to derive the clock from the stream of received packets must contend with ambiguous transitions, whereas a synchronised member of the Piconet can use the known clock to assist in processing the RF stream. This problem is even more compounded in later Bluetooth versions, with v2.0 introducing Enhanced Data Rate (EDR), and v3.0 adding High Speed (HS), also referred to as “Alternative MAC/PHY” (AMP). EDR makes use of different RF modulation, depending on the packet type being sent. For the majority of link management purposes, the previously defined GFSK modulation is used; however, for data packets, particularly those involved in the delivery of audio services, a more complex Phase Shift Keying (PSK) modulation is used \cite{BluetoothSIG2007} . 

When this scheme is in use, the modulation applied in a specific communication session will change frequently based on the data being sent. The sniffer’s challenge of discriminating between spurious radio signals and actual data becomes markedly harder, as confirmed by \cite{Naggs2013}.

\begin{comment}
Bluetooth 3.0’s AMP/HS takes this concept further, with a completely separate PHY/MAC stack being used for bulk data, with the Bluetooth channel only being used for control. Access to the Radio Spectrum is controlled using a Time Division Duplex/Time Division Multiple Access scheme. All devices in the net use the same RF channels (“Multiple Access”) but only one device can speak at a time (“Time Division”) and take turns in transmitting and receiving (“Duplex”). 
\end{comment}

Bluetooth is designed to support a hierarchy of Piconets and Scatternets , however, this usage has not been adopted widely, and in practice, most Bluetooth communication is between a single master  and single slave device, such as a smartphone and car. In this scenario, the master device will transmit on even numbered hops, whilst the slave device will transmit on odd numbered hops \cite{Chen2012}.

Pelzl and Wollinger’s other contribution is to describe a series of limitations that they identify in Bluetooth’s security. This 2006 list is largely an adaptation of \cite{Jacobsson2001}, however, they make the definitive statement "It is possible to intercept radio signals originating from Bluetooth devices (e.g. with a Bluetooth protocol analyzer ...)".  While this is an enduring notion amongst researchers, the work performed by \cite{Spill2007}, \cite{Ossmann2009} and \cite{albazrqaoe2016practical} has demonstrated that practically intercepting traffic – particularly in a passive fashion – is a significantly harder task than these early authors had anticipated. 

\subsubsection{Basic Hopping Sequence}
Bluetooth uses a Frequency Hopping Spread Spectrum (FHSS) mechanism. 
The ISM band from 2402MHz to 2480MHz is divided into 79 channels of 1MHz each. The edges of the band (2400-2402MHz and 2480-2483.5MHz) are not used. 
\cite{Pelzl2006} predates the development of Bluetooth Low Energy (BTLE), which uses a different channel division schema, separating the same RF spectrum into 40 channels, with 2MHz of bandwidth each \cite{Ryan2013}. The two schemas are compatible at the RF layer, and can interoperate in the same physical space, but require different mechanisms of link control. 

Bluetooth devices maintain an internal 28-bit 3200Hz clock. During normal communication, the upper 27 bits of the clock, $Clock_{27}$ is used and each of the 79 available channels are used for only $625 \mu s$ before communication \emph{hops} to the next channel in the sequence; this means there are 1,600 \emph{slots} per second, and the clock increments twice for each time slot \cite{Spill2007}. The hopping sequence is not random – it is pseudo random, calculated using the \emph{Hop Selection Kernel}; an algorithm defined in the v1.1 Core Specification and modelled in detail by \cite{Albazrqaoe2011}. The kernel is seeded with the following values:

\begin{itemize}
\item The UAP and LAP of the master device; and
\item Bits 1-26 of the clock index.
\end{itemize}

\begin{figure}
  \includegraphics[width=\linewidth]{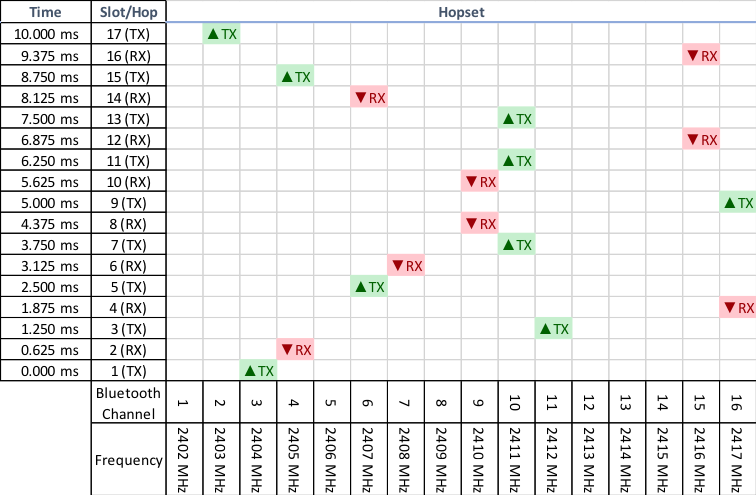}
  \caption{Simplified Hopset showing 16 Channels, 10ms of Hops, 2 Devices}
  \label{fig2}
\end{figure}

These are combined to define the RF Channel index, the next channel to be hopped to.  Figure \ref{fig2} shows a simplified example, with only 16 channels, demonstrating how hops proceed during normal communication, from the perspective of the master device. In this limited hopset, on each successive “slot”, the master device will transmit the data which it wishes to send (if any), then wait for a response on the next slot. This re-iterates an important behaviour; Bluetooth uses TDMA “time division multiplexed access” to determine when it can transmit or not, based on these rotating time slots, rather than the CSMA/CA “carrier sense multiple access with collision avoidance” used in IEEE 802.x wireless standards \cite{Pelzl2006}. 

\subsubsection{Adaptive Hopping Sequence}
Adaptive Frequency Hopping was introduced in Bluetooth version 1.2, ratified in 2003. This approach improves resilience in an environment where Wi-Fi or other ISM technologies are being used – those channels which cannot be reliably used because of interference from Wi-Fi users and access points are marked as “bad” and the usable channels available for frequency hopping, the \emph{hop set}, is reduced accordingly \cite{Hodgdon2003}.

As described by \cite{Popovski2006}, the behaviour of Frequency Hopping systems is governed by rules laid down by the regulatory bodies who control access to the radio spectrum, such as the FCC. To comply with these regulations, frequency hopping must continue at high enough a rate to ensure that the \emph{dwell time} on any given channel is no longer than 0.4 seconds in a given hop. Bluetooth supports a minimum hop set of 20 channels, and is able to retain the same hopping rate, 1600 per second. 

An AFH Channel Map is maintained by the master device of the piconet – a 79 value table where each channel is marked as “good”, “bad” or “unknown”. The table is sent from the master device to all slaves in the piconet using the Link Management Protocol (LMP) command LMP\_Set\_AFH() \cite{BluetoothSIG2007}. Slave devices can ask the master to exclude a channel, however, the master makes the decision, and communicates the updated map each time it is changed. It should be noted that the AFH scheme is dynamic, and channels can be added to the hop set again, as well as removed. 

\begin{figure}
  \includegraphics[width=\linewidth]{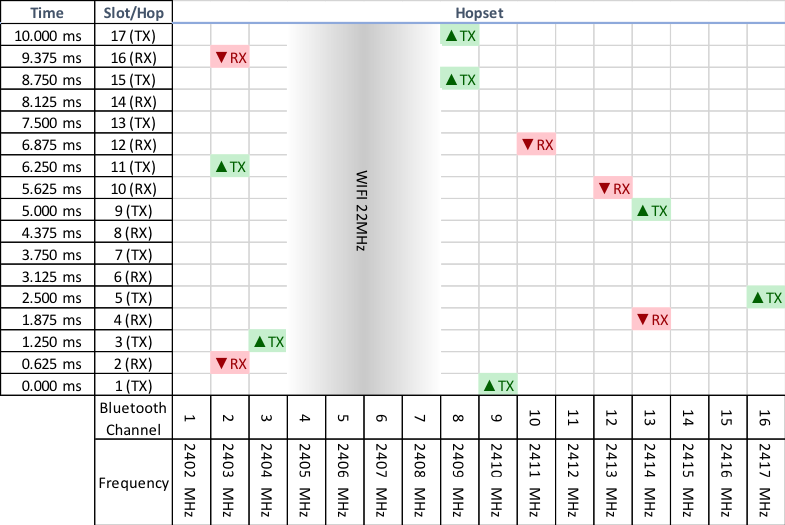}
  \caption{Simplified Hopset with AFH mitigation of Wi-Fi interference, 2 Devices}
  \label{fig3}
\end{figure}

Figure \ref{fig3} shows a simplified hopping scheme using only 16 channels. In practice, all 79 channels are available, and a Wi-Fi channel can obstruct as much as 22MHz of available bandwidth; as many as 11 Bluetooth channels above and below the Wi-Fi channels nominal centre frequency. 

\begin{figure}
  \includegraphics[width=\linewidth]{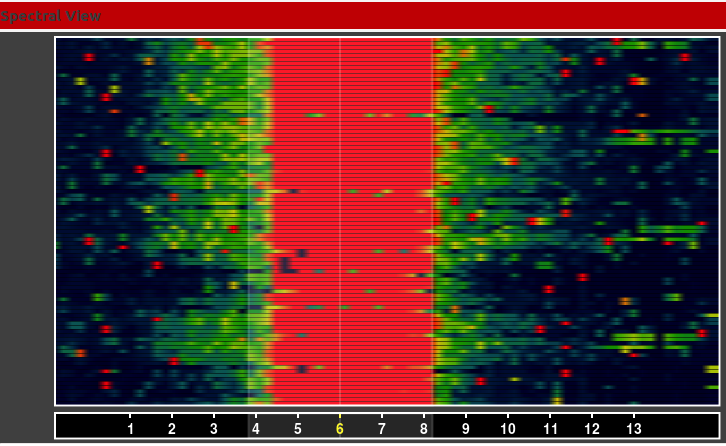}
  \caption{Spectrum Analysis - RF usage by Busy Wi-Fi}
  \label{fig4}
\end{figure}

This can be seen in Figure \ref{fig4} – a capture of the ISM band using the experimental setup described later, which shows the broad footprint of Wi-Fi Channel 6 in heavy use. 
 
The X-axis represents the entire ISM Band, from channel 0 at the origin to channel 79 at the right-hand edge. The Y-axis represents time - new samples are added at the top, and the display scrolls downwards, removing the oldest sample from the bottom. Each cell represents an RSSI (Received Signal Strength Indicator), visualising the signal strength as a “heat” map. The scale is “decibel milliwatts” or dBm, where blue/black represents a weak signal of -90dBm or less, continuing through green and yellow until red, which represents the strongest signals of -60dBm. 

\begin{comment}
Received Signal Strength is measured relative to the transmitted signal, so is expressed as a negative number, further, dBm is a logarithmic measure, so a reduction of 10dB in RSSI represents a signal that is ten times stronger. 

In addition to the RF spectrum occupied by Wi-Fi, this chart also shows Bluetooth’s hopping behaviour – the red cells distributed in an apparently random manner throughout the spectrum are caused by Bluetooth audio streaming to a set of speakers. 
\end{comment}

As for Basic hopping, the choice of next frequency to hop to is based on the Hop Selection Kernel. From version 2.1 onwards, this has been extended to support AFH, and is now seeded with the following:

\begin{itemize}
\item The UAP and LAP of the master device;
\item Bits 1-26 of the clock index;
\item The AFH Map supplied by the master device; and
\item n, a numeric value. This is the number of usable channels in the hop set. 
\end{itemize}

The reduced hop set in this scenario means that any available channel is selected more frequently by the hopping algorithm than in an environment where AFH is not used.  This structure was hypothesised by Spill \cite{Spill2007} and again by Huang \cite{Huang2014} to create a potential advantage to a would-be sniffer because the likelihood of detecting traffic for a given piconet increases whilst listening on a single channel. 

\begin{comment}
In practice, the process by which a Bluetooth master device decides that channels are “good” or “bad” is explicitly not defined by the SIG specification \cite{BluetoothSIG2007} and varies from vendor to vendor in a non-transparent fashion, making it less advantageous to a would-be sniffer.  
\end{comment}

\subsubsection{Finding and Joining a Piconet}
Piconet link management relies on two mechanisms; Inquiry, which is intended for discovering new devices and establishing the required information to join the piconet and Paging, which is intended to allow a device to join a piconet and begin the pairing process \cite{Tabassam2007}. In both modes, a smaller set of 32 evenly distributed “wake-up” frequencies are used across the same 79MHz band as the basic and adapted hopping channels \cite{BluetoothSIG2007}.

\begin{comment}
Confusingly, the terminology for inquiry and paging is different from the rest of the Bluetooth specification. The term \emph{master} is used, not to describe the controlling node in a piconet, but the device initiating communications. Similarly, ‘slave’ is used, not to describe a synchronised participant in a piconet, but rather the device responding to inquiries. In the following section, we use the terms ‘inquiring device’ and ‘listening device’ for clarity. 

There are two key differences which support rapid discovery. Firstly, 

\end{comment}Unlike the basic hopping channel, where master and slave devices hop in a synchronised fashion, in Inquiry mode, the inquiring device hops on each tick of $Clock_{0}$ - 3200 times per second, rather than the 1600 hops per second of $Clock_{1}$. On each slot, it transmits an ID packet containing an Inquiry Access Code (IAC).

\begin{figure}
  \includegraphics[width=\linewidth]{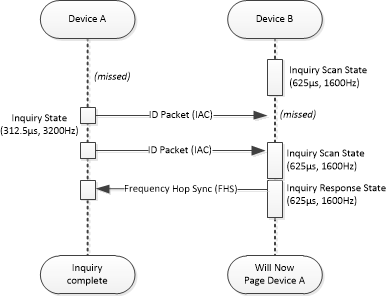}
  \caption{Inquiry Process (adapted from (\cite{Tabassam2007})}
  \label{fig5}
\end{figure}

Bluetooth devices in discoverable mode enter an \emph{inquiry scan state} (Figure \ref{fig5}), where they listen for IAC messages at least once every 1.28 seconds, for at least 11.25ms. During this period, the listening device continues to hop and lingers on the designated \emph{wake up} frequencies according to the usual 1600Hz pattern.

Secondly, the inquiry method allows the use of the same frequency for communication back from the discovered device. This means that a device which is scanning to join a piconet is more likely to coincide on the same channel as the listening device, and able to respond without waiting for another collision. 
The listening device responds by sending a specific Frequency Hop Synchronisation (FHS) packet which contains its 48-bit BD\_ADDR, and the current clock, $Clock_{27-1}$.

It is important to acknowledge that the only difference between a \emph{discoverable} and \emph{non-discoverable} Bluetooth device is whether it responds to inquiry. An attacker who knows the information supplied in the FHS packet can proceed directly to the Paging process. In either case, the inquiring device is now able to send a Page Request to the device to initiate a connection. Whilst the previously inquiring device now has frequency hopping information, and could align with the basic or adaptive hopping channels, the Paging process is also completed using the 32-channel set of wake-up frequencies, and at the accelerated clock rate of 3200Hz.

\begin{figure}
  \includegraphics[width=\linewidth]{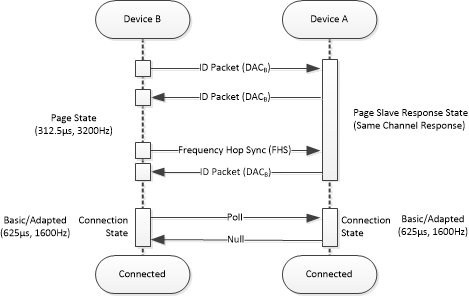}
  \caption{Paging Process (adapted from \cite{Tabassam2007})}
  \label{fig6}
\end{figure}

Note that the perspective has now changed – Device B is initiating communications, and Device A is responding. On each timeslot, Device B sends another ID packet, this time directed to Device A’s BD\_ADDR, and containing its own Device Access Code (DACB), as illustrated in Figure \ref{fig6}.

Device A responds by echoing the DAC, at which point the Device B will send an FHS packet containing the piconet information (BD\_ADDR, $Clock_{27}$ and AFH Map if needed).  Device A sends another echo of the DAC as an acknowledgement, and both devices now hop to the AFH\_Instant specified in the FHS message. At this point, the devices are connected and the inquiring device is now able to participate in the normal hopping sequence of the piconet. A Poll/Null ping is used to confirm that the connection is correctly established. 
 
Any Bluetooth device in a connectable mode (that is, any normally operating device) monitors the page scan channel, and will respond to a page in this fashion. This is why researchers such as Spill \cite{Spill2007} indicate that making a device non-discoverable is not a defence against exploits. It is enough to know a device's BD\_ADDR and local clock to connect \cite{Scarfone2008}, which is of particular relevance given the findings of Seri \cite{Seri2017}. In describing the BlueBorne vulnerability set, they highlight that the BD\_ADDR of most smartphone targets can be easily recovered by sniffing of Wi-Fi traffic.

\subsection{Addressing and Network Formation Terminology}
\begin{figure*}
  \includegraphics[width=\linewidth]{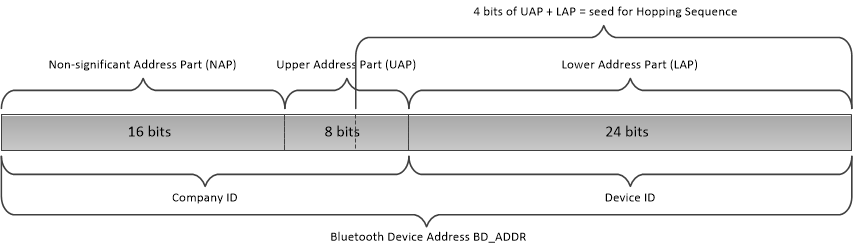}
  \caption{Bluetooth Address Parts}
  \label{fig7}
\end{figure*}

\begin{figure*}
  \includegraphics[width=\linewidth]{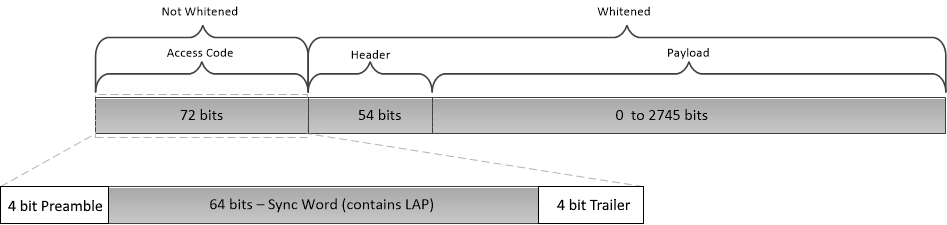}
  \caption{Bluetooth Generic Packet Structure, header detail including LAP}
  \label{fig8}
\end{figure*}

Each Bluetooth device should have  a globally unique address (BD\_ADDR), which uses a similar format to the IEEE 802.x MAC address used by Ethernet devices. This is a 48-bit number, issued under IANA rules by approved vendors, and typically represented as a series of Hexadecimal digits. Device manufacturers are issued a 24-bit value, which represents a range which they have authority over, e.g. \emph{00a0c6} has been issued to Qualcomm for the devices which they manufacture (Figure \ref{fig7}). Significantly, whereas the entire IEEE MAC address is transmitted and received in other 802.x family protocols, in Bluetooth, only the LAP is transmitted over the air, as part of the packet header, as shown in Figure \ref{fig8}.
 
The UAP does, however, play a critical role in RF communications; the lower four bits of the UAP are combined with the 24 bits of the LAP, to produce the address value used by the hopping sequence algorithm. The NAP is, as the name suggests, non-significant; it is not used by the Bluetooth protocol in any way – Spill \cite{Spill2012} confirms this by demonstrating the NAP portion set to 00:00, confirming that this has no effect on operation. 

\begin{comment}
Bluetooth describes two forms of network formation – the Piconet and Scatternet. The Piconet consists of a single master device, and from one to seven active slave devices (Figure \ref{fig9}).  A piconet can only have one master device, but a slave device can participate in more than one piconet; this extended network is referred to as a scatternet. Direct communication between slave devices is not supported. All communication is between slaves and master devices. Additionally, no routing of data between piconets is described within the Bluetooth standard up to v5.0. 

Despite being present in the standard since v1.0, scatternet functionality has not seen wide adoption, and the majority of Bluetooth usage is a simple piconet with only two devices \cite{Chen2012}. The proposed next version of the standard v6.0 introduces Bluetooth Mesh, which seeks to enhance scatternets with mesh routing capability and provide a degree of slave to slave communication, in order to deliver functionality more in line with Zigbee, a competing ISM-based radio protocol \cite{OMalley2017}.

\begin{figure}
  \includegraphics[width=\linewidth]{figures/image009.png}
  \caption{Topology of example Piconets and a Scatternet}
  \label{fig9}
\end{figure}  
\end{comment}

\subsection{Pairing and Authentication}
Prior to v2.1 of the standard, pairing encryption and authentication used the Ex series algorithms, based on SAFER and SAFER+ (Table \ref{tab:2}). With successive updates to the standard, more secure mechanisms were introduced in response to criticism and the hypothesised weaknesses described by Jacobsson and Wetzell \cite{Jacobsson2001}, namely that the link key was recoverable, and the cipher weakened due to poor key management.

\begin{table*}[t]
  \caption{Evolution of Bluetooth Cipher Suite (adapted from \cite{Scarfone2008})}
 \begin{tabular}{p{3cm} | p{3cm} | p{6cm} | p{3cm}}
Description &	Prior to 2.1 &	2.1 - 4.0 &	4.1 Onwards \\
\hline 
Pairing Algorithms	& $E_21$, $E_22$, SAFER+ & P192 Elliptic Curve,HMAC-SHA-256 (Secure Simple Pairing), $E_21$, $E_22$, SAFER+  (Legacy Pairing)	 & P256 Elliptic Curve, HMAC-SHA-256\\
Encryption Algorithm & $E_3$ / $E_0$ / SAFER+ & $E_3$ / $E_0$ / SAFER+ &	AES-CCM\\
Device Authentication & $E_3$ / $E_1$ / SAFER & $E_3$ / $E_1$ / SAFER		 & HMAC-SHA-256\\
\end{tabular}
  \label{tab:2}
\end{table*}

It should be noted, however, that backwards compatibility has been retained by the SIG, and this means that even in the latest v5.0 standard, legacy $E_0$ encryption, and pairing using the Ex series algorithms is still supported – indeed, if all devices in a piconet cannot support AES-CCM encryption or HMAC-SHA-256 authentication, then all devices on the piconet will downgrade to the legacy mechanisms \cite{BluetoothSIG2016}. We therefore examine the legacy pairing and authentication schemes as an example.

\subsubsection{Legacy Bluetooth Pairing and $E_1$ Authentication}
The legacy sequence takes seven packets \cite{Shaked2005} (Table \ref{tab:3}).

\begin{table*}[t]
  \caption{Seven packets required to perform Shaked and Wool \cite{Shaked2005} PIN attack}
 \begin{tabular}{p{1cm} | p{1cm} | p{1cm} | p{3cm}  | p{3cm}  | p{3cm}}
Packet	& Src	& Dst	& Data	& Length &	Notes \\
\hline 
1 &	A &	B &	IN\_RAND &	128 bit &	Plaintext\\
2 &	A &	B &	LK\_RANDA &	128 bit	&  XORed with $K_{init}$\\
3 &	B &	A &	LK\_RANDB &	128 bit	& XORed with $K_{init}$\\
4 &	A &	B &	AU\_RANDA &	128 bit	& plaintext\\
5 &	B &	A &	SRES &	32 bit &	plaintext\\
6 &	B &	A &	AU RANDB &	128 bit &	plaintext\\
7 &	A &	B &	SRES &	32 bit &	plaintext\\
\end{tabular}
  \label{tab:3}
\end{table*}

\begin{figure}
  \includegraphics[width=\linewidth]{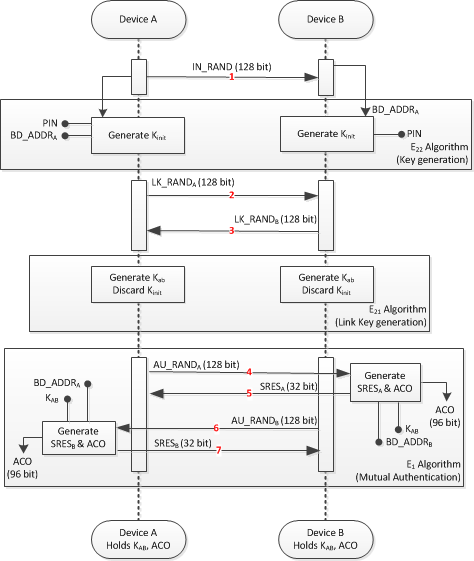}
  \caption{Key Generation and Authentication using Ex Series Algorithms}
  \label{fig10}
\end{figure}

This sequence is shown in Figure \ref{fig10}, where Device A is the initiator and Device B is the responder. After the first packet, each device holds the IN\_RAND initialisation value, and each knows the PIN by other means, typically being entered on the device by the user, or in the case of simpler devices, hard-coded to a known value such as 0000 or 1234. Each device generates an initialisation key $K_{init}$, generates another random number, xORs this with their $K_{init}$ and passes this to the other device. Each device uses these values, combined with their own $K_{init}$, to generate the Link Key $K_{AB}$. This key persists for the duration of the pairing.

Finally, each device produces another random number. The device uses this random number, the BD\_ADDR, and the Link Key $K_{AB}$ to generate another 128-bit number – the top 32-bits are $S_{RES}$ and the lower 96-bits are the Authenticated Ciphering Offset (ACO) \cite{Scarfone2008}. The device then passes its random number to the other device. When each device has verified that its own calculated value of $S_{RES}$ matches the value returned by the other device, each knows that the other device holds a valid copy of the Link Key, and has derived the same ACO for the link \cite{Chen2012}.

\subsubsection{Data Whitening}
\begin{figure}
  \includegraphics[width=\linewidth]{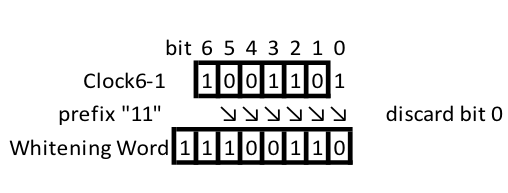}
  \caption{Generation of Whitening Word from $Clock_{6}$}
  \label{fig16}
\end{figure}

Prior to assembling a Bluetooth packet, the 54-bit header and payload are \emph{whitened}. The whitening process, which is reversed at the receiver as \emph{de-whitening}, involves passing the data through a Linear Feedback Shift Register (LFSR) which is preloaded with a whitening word. The initial word is derived from $Clock_{6}$ which is transposed as shown in Figure \ref{fig16}.

The LFSR changes state in a predictable fashion with each operation, and XORs the current value with each bit in the data to be transmitted in turn. The process is repeated at the receiver, and, as long as the LFSR is initialised with the same whitening word as used to scramble the packet, allows the data to be unscrambled in the same fashion.

Whitening is not technically part of the encryption process, as it is not performed to obfuscate the data, but primarily to remove any long chains of zeros or ones to assist the performance of the analogue electronics in the RF stage, and prevent DC Bias issues \cite{BluetoothSIG2007}. The process does, however, add an extra degree of complexity in recovering a packet from the radio transmission \cite{Spill2012}.

\subsection{The Development of Hypothetical Attacks on Bluetooth}

From version 1.0 of the Protocol onwards, attack methods have been proposed, such as Jacobsson and Wetzell \cite{Jacobsson2001}, who highlighted flaws which they believed to be significant in the Bluetooth specification. 
\begin{comment}
\begin{enumerate}
\item A weakness in the pairing process which allowed the PIN used to determine the Link Key to be discovered;
\item The potential to use the Bluetooth device’s ease of pairing to allow the user’s location to be determined; and
\item A weakness in the encryption scheme used once a link has been established.
\item The potential to use Bluetooth device emissions for location detection and mapping of individual human actors remains a possibility, however, these concerns appear to have become culturally less significant, given the far greater usage of location services and human interaction tracking which has become \emph{normal} for smartphone users in the intervening 16 years. 
\end{enumerate}

Jacobsson and Wetzell \cite{Jacobsson2001} observe that other researchers had raised concerns that the $E_0$ cipher used in the Bluetooth 1.0 specification was prone to attacks, and was weaker than the other elements of the scheme (E1, $E_21$, $E_22$) which were based on equivalent elements of the well understood SAFER+ crypto scheme.

The issue raised which has most relevance to the development of sniffing attacks is the proposed ability to perform a Man in the Middle (MITM) attack, and to determine the PIN used to generate the link key as a listening third party. 
\end{comment}
The authors assert that the inherent difficulty of frequency hopping cannot be relied on to provide security, however, their assertions about the ease of overcoming this difficulty do not appear to be well supported.  In short, these authors describe a theoretical weakness but did not demonstrate a practical means by which the weakness could be exploited\cite{Jacobsson2001}.

Shaked and Wool \cite{Shaked2005} laid out the fundamentals of such an attack more clearly. The pairing sequence takes seven packets. Packet 1 contains IN\_RAND, the initialisation value used to generate $K_{init}$. A would-be attacker can use this value, BD\_ADDRA and repeat the $E_22$ algorithm with a guess for the PIN to generate a possible value of $K_{init}$. As the PIN is a 4-6 digit number, this is possible to brute force offline in a trivial amount of time. Shaked and Wool calculated that a high specification computer of the day (a Pentium 4 3.0GHz based machine) could brute force a 4-digit pin within 0.063s, whilst a 6 digit pin could be recovered within 7.26s\cite{Shaked2005}. A current high spec PC based on the Intel Core i9XE processor could brute force the larger six digit pin within 0.096s; effectively instantly in practical terms \cite{CPUBenchmark2017}. 

Packets 2 and 3 contain LK\_RANDA and LK\_RANDB – random 128-bit values chosen by each device, XORed with $K_{init}$ – the postulated value for $K_{init}$ is used to retrieve these, and this set of information is now enough for the attacker to use $E_21$ to guess the Link Key $K_{AB}$.

The $E_1$ algorithm is then used with the guessed Link Key to perform the mutual authentication process with the AU\_RANDA and AU\_RANDB retrieved from packets 4 and 6. If the computed SRESA and SRESB values are correct, then the attacker holds a valid $K_{AB}$, and by definition, has guessed the correct pin. If not, another PIN is chosen and the process repeated. 

The authors acknowledge that to actually perform the hypothesised attack, it is necessary to \emph{Assume that the attacker eavesdropped on an entire pairing and authentication process and saved all of the messages} \cite{Shaked2005}.

Bluetooth Packets can take 1, 3 or 5 timeslots \cite{Rivertz2005}, however each of the 7 pairing packets required by the author’s method are short – containing only 32 or 128 bits of data respectively. Each packet in the pairing sequence therefore takes only a single timeslot. 

It is not possible to hop along with the devices as they pair without either completing the inquiry/paging process, which renders the attack active, rather than simply eavesdropping, or by deducing the BD\_ADDR of the master device, and the $Clock_{n}$ value. If an attacker is listening on a single channel, therefore, there is only a 7 in 79 chance (8.86\%) that any of the packets involved in the pairing sequence can be observed, and an absolute certainty that the full exchange will not be seen.

The simple statement by Shaked and Wool \cite{Shaked2005} of  \emph{assume that...}, reflects the ease of carrying out such capture in other 802.x protocols. An assumption is made that capturing Bluetooth data is similarly easy, whilst attacking the protocol or crypto elements is harder, therefore if a weakness in these can be demonstrated, that it will be trivial to exploit in practice. Indeed, they go so far as to advise against forcing devices to freshly pair each time they communicate, instead storing link keys and using Bluetooth’s re-pairing functionality to avoid exposure to the weakness they describe. The lack of practical evaluation of this, and other, hypothesised attacks appears to be a gap that could be addressed experimentally.

\subsubsection{BlueSniff}
Spill and Bittau \cite{Spill2007} published a paper, BlueSniff.\begin{comment}, which is derived from and expands upon Spill’s presentation at Usenix’s first Workshop on Offensive Technologies (WOOT) in 2007.\end{comment} They suggested that, despite the attacks demonstrated against the upper layers, the underlying Bluetooth protocol remained relatively secure – primarily due to the practical difficulties in eavesdropping packets sent between devices. 

Spill used a Software Defined Radio (SDR) platform called the Universal Software Radio Peripheral (USRP).
\begin{comment}, and spent a considerable time understanding how to capture and demodulate the radio signal. He notes that the firmware of Bluetooth chipsets is not open sourced, and that simply tuning a radio to the correct frequency is not enough. 
\end{comment}
The USRP platform proves limiting for two reasons – firstly, it is expensive at over \$2000 USD, and secondly, it was not intended for FHSS use. In the 2.4GHz range, the USRP takes $200\mu s$ to stabilise on a given frequency; given that an entire Bluetooth timeslot is $625\mu s$ in basic hopping and only $312.5\mu s$ in page/inquiry scan mode, this means that it is not capable of participating in frequency hopping. 

The BlueSniff paper therefore describes the approach taken by the authors to monitor multiple channels without honouring the hopping behaviour – greatly assisted by the discovery of a debug mode in Cambridge Silicon Radio’s (CSR) Bluetooth development kit which forces packets to be broadcast on a single channel \cite{Spill2007}.

Spill highlights three specific issues preventing eavesdropping – the lack of an integral ‘promiscuous’ mode, as is offered in the PHY of other 802.x protocols, the difficulty posed by the scrambling or ‘whitening’ of the data, and finally, the requirement to recover the master's BD\_ADDR. 

Whitening is the most significant of these for Spill, and means that it is not possible to use techniques such as transmitting a block of data, then looking for that specific pattern in the captured radio transmission. To do this, the packets need to be extracted, and de-whitened. The whitening of Bluetooth packet elements is based on a whitening word derived from the master device’s clock index, $Clock_{6}$. This is transposed and a two byte prefix added as shown in Figure \ref{fig16}. As the first 2 bits are always “11”, the remaining 6 bits of the whitening word present only 26 = 64 possible starting values for the LFSR. 

Spill's approach, and the first significant contribution of the paper, is to use each of the 64 possible starting positions for the LFSR, to perform the de-whitening process, generate each of the 64 possible packets, then recalculate the Cyclic Redundancy Check (CRC) of the payload. If the CRC matches, then it is likely that the proposed whitening word is correct, and from this, the $Clock_{6}$ can be recovered. A limitation which the authors recognise is that not all Bluetooth packets require a CRC on the payload. In practice, multiple packets may need to be analysed before a suitable candidate is found. With a means to de-whiten packets, Spill focuses on recovering the BD\_ADDR. The LAP is easily recovered, as it is included in the header of every packet transmitted on a piconet in the 72-bit access code.

Spill’s second significant contribution is a means to recover the UAP, which is not included in the transmitted packet. As a radio protocol designed to be tolerant of noisy, congested environments, Bluetooth makes extensive use of error correction which proves to be the key. Each packet’s 54-bit header includes a Header Error Check (HEC) value. This HEC is generated and checked using a similar LFSR method to the whitening process, however, it is initialised at both ends of the link using the 8-bit UAP, taken from the master device’s BD\_ADDR \cite{BluetoothSIG2007}. Spill recognised that as a fundamentally bidirectional XOR process, the LFSR can be loaded with the HEC value as the initialisation word, then the header can be replayed in reverse bitwise. At the end of this process, the LFSR will contain the previous initialisation value, that is, the UAP. The paper represents a significant breakthrough, but the authors are careful to highlight the weaknesses of their approach. In considering future directions, Spill and Bittau postulated that restricting the channels available for Bluetooth to use through manipulation of the AFH map might provide a means to narrow the attack – it is this recommendation which is investigated further by this paper. They also discussed using multiple USRP’s, each monitoring five channels to capture multiple channels at once. 

Spill \& Bittau recognised that to calculate the hopping schedule, the full clock would be needed rather than simply the clock offset used in whitening, and proposed that, with the BD\_ADDR recovered, it should be possible to connect to the piconet master using paging mode and recover the clock. This would, of course, represent an active attack, rather than passive eavesdropping. 

\subsubsection{The Ubertooth Project}
\begin{figure}
  \includegraphics[width=\linewidth]{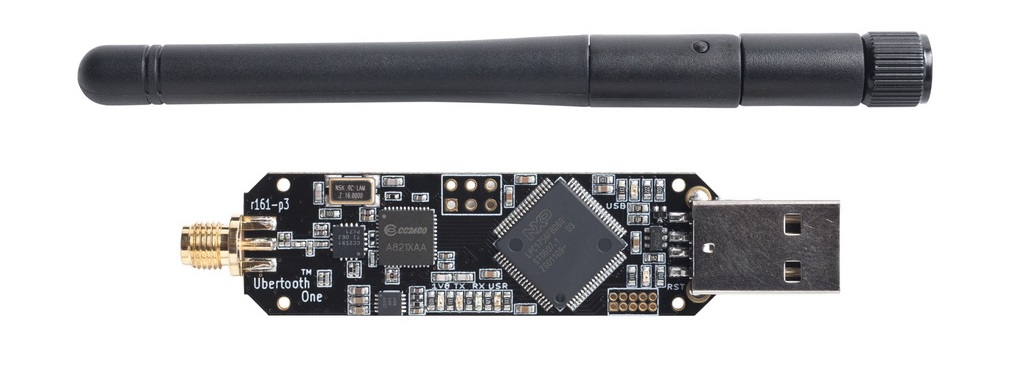}
  \caption{Ubertooth One adapter with SMA antenna}
  \label{fig17}
\end{figure}

The development of Ubertooth was discussed during the presentation by \cite{Ossmann2009} at ShmooCon in 2009. In an attempt to address the issues of the USRP platform and provide a lower cost tool for researchers, Ubertooth Zero was developed and made available as an open sourced hardware design. This tool built on the proposed research outlined in \cite{Spill2007}, specifically, the ability to follow Bluetooth’s hopping sequence and perform some elements of the brute forcing required in silicon rather than code. In his presentation to RUXCON \cite{Spill2012} introduced Ubertooth One (Figure \ref{fig17}), and described the research of the Ubertooth project to date. 

Spill describes the Ubertooth One device, but also highlights a breakthrough in thinking about clock recovery – a mechanism to recover the full clock from traffic without the need to connect to the piconet as required by the method proposed in his 2007 paper.  For a known UAP + LAP, all 227 hops in the basic hopping sequence before the clock wraps around can be calculated. The clock index cannot be determined by observing a single packet; however, as the actual hopping sequence is observed, the pattern can be compared to the predicted sequence to find a match. 

When a sequence of observed hops matches the predicted sequence, a prediction of the next hop can be made, and used to confirm the guess. This provides the value of the clock index.  Once the clock index is known, the regular $625\mu s$ slots can be used to increment the clock used by the eavesdropper. 

\begin{comment}
Spill highlighted that this process could be accelerated through the use of ubertooth-follow, a tool which makes use of a second Bluetooth dongle participating legitimately in the piconet. Again, this is an active attack, rather than passive eavesdropping, however, this approach demonstrates the first use of a second Bluetooth device as a side channel to support an attack. Despite his contribution, \cite{Spill2013} describes the gap between perception – that Bluetooth is as simple to sniff as other wireless protocols, such as Zigbee and Wi-Fi – and the reality, that \emph{sniffing Bluetooth is hard}. 

As recently as 2016, despite ongoing investigation, the Ubertooth team contend that it is still not possible to use consumer purchasable hardware to sniff the Bluetooth traffic between two devices which are not previously known to the attacker \cite{Spill2016}. The work has, however, been carried forward by other researchers.
\end{comment}

\subsubsection{BlueID}
The BlueID paper \cite{Huang2014} describes a means to fingerprint and subsequently identify specific Bluetooth devices, without attempting to follow the frequency hopping sequence, or understand any of the higher protocol elements. The paper is the first collective publication of this team of researchers at Michigan State University (MSU) and builds upon the exploration of the frequency hopping mechanism by \cite{Albazrqaoe2011} in his Masters thesis, itself building on \cite{Ossmann2009}.\begin{comment}
 The clock of the slave devices in a piconet needs to be synchronised to that of the master; the Bluetooth specification allows for $ \pm  10\mu s$ drift for the start of each timeslot. 

The authors use the \emph{jitter} that is inherent in analogue RF electronics to identify and measure this drift of clocks which proves consistent enough to be stable, and unique enough for any given device to provide a usable fingerprint.
This work is restricted, in that it only samples a single channel out of the 79 used, and only ever uses the BD\_ADDR element to confirm that a given fingerprinted device is, indeed, valid. It is significant, however, for the development of thinking that this demonstrates in the MSU team. This understanding of clock drift and the effect of interference on selection (or deselection) of channels for the AFH map has played a significant role in the development of the subsequent BlueEar project \cite{albazrqaoe2016practical}.
\end{comment}

\subsubsection{BlueEar}
In their BlueEar paper, Albazrqaoe et al \cite{albazrqaoe2016practical} consider the use of dedicated Bluetooth sniffing equipment capable of listening to all 79 channels at once, and propose a low-cost platform using two Ubertooth One adapters; one as a “scout” and the second as a “sniffer”. Again, they highlight that a potential eavesdropper cannot follow the hop sequence of a piconet unless they know the BD\_ADDR of the master, and the current clock index. 

They describe brute force clock acquisition, which is similar to Spill and Bittau’s hypothesised approach; however, they perform the work of extending the mechanism to consider the effects of AFH. \begin{comment}
 – initially explaining how the remapping of bad channels by AFH means that even when the correct clock has been guessed, the observed hopping behaviour may not match the prediction. FCC rules require a minimum of 20 channels to be used for FHSS, even in heavily congested environments. The MSU team use this information to weight the likelihood of a guessed clock index being correct; if the ratio of \emph{misses} (where the observed behaviour does not match the prediction) exceeds (79-20)=59/79, then the guessed clock index may be considered as wrong.
\end{comment}

They make two further significant contributions. Firstly, they observe that Bluetooth Classic in AFH mode will only transmit on channels which the master device considers “good”. This observation is used to make two further deductions; if the packet rate observed on a given channel is in the Top 20, then those channels are likely deemed by the master as \emph{good}. Further, given the FCC rule, the average packet rate of these Top 20 can be used as a good approximation for the packet rate of the piconet as a whole. Conversely, knowing the average packet rate of the piconet as a whole, it is reasonable to assume that the channels which have packet rates significantly below this average value are considered by the master of the piconet as \emph{bad}.

Secondly, they consider another implication of the master device’s behaviour in selecting \emph{good} and \emph{bad} channels; that the master will consider a given channel to be bad if it is subject to interference from other devices or ISM users. They make use of their two radio solution by using one of the radios to hop between all 79 channels measuring the apparent noise level on that channel.

Where a channel is particularly noisy, it is reasonable to assume that the piconet master will consider this channel to be \emph{bad} and exclude it from the hop set. These additional items of information help a prospective sniffer to build their own replica of the AFH map held by the master device – in turn, this allows for more accurate prediction of the hopping sequence. \cite{albazrqaoe2016practical} have developed their approach across a series of papers, and have carried forward Spill and Ossmann’s ideas significantly.

\section{Methodology}
\subsection{Introduction}
Spill, Albazrqaoe and Checkoway in their respective papers quantify the extent to which deliberately congesting the ISM band can force real world Bluetooth AFH implementations to abandon the congested frequencies, and whether the adoption of this updated AFH Map has a quantifiable, measurable effect on the success rate of packet capture. There are a variety of well understood mechanisms to accomplish this, such as RF noise generators or Wi-Fi jamming devices, however, these appear to be of questionable legality and limited availability. Instead, an approach was sought using a commonly available, legal to use, technology – consumer Wi-Fi devices. 

\subsection{Experimental Method}
As a starting point, the experiments will use the technique described in BlueSniff \cite{Spill2007}. Subsequent researchers \cite{Huang2014}, \cite{albazrqaoe2016practical} and \cite{Chernyshev2017} had each chosen the Ubertooth hardware developed by Spill and Ossman, and have used the supporting software tools to examine similar research questions. The literature did not describe any superior alternative mechanisms, so Ubertooth was chosen as a platform.

The planned investigation is to measure the effectiveness of manipulating the AFH Map by congesting the ISM band in improving the ability to sniff data passively. Examining this step by step, to measure the effect on sniffing data passively, a metric must be identified which can be used as a benchmark to compare one capture attempt to the next. A means must be developed of capturing the AFH Map in effect, and a mechanism to determine the degree of congestion of the ISM band understood. 

Additionally, the experiment should allow these measurements to be made in a repeatable fashion, and be carried out sufficient times to allow a reasonable body of data to be gathered.
Therefore, the experimental method must have the following characteristics:

\begin{itemize}
\item A means to compare one capture run to another in qualitative terms;
\item Only a single aspect of behaviour should be measured in each experimental setup;
\item A mechanism to ensure that each run is different only in terms of the aspect being investigated and any external factor should be controlled as far as possible;
\item The experiments should be repeatable; and
\item The experiments should be repeated to allow a meaningful amount of data to be gathered.
\end{itemize}

The process for capturing data described in \cite{Spill2012} involves the brute forcing of the $Clock_{27}$ as a precursor to decoding packets. At the simplest level, failure to acquire $Clock_{27}$ means that no data can be captured, whilst a rapid, early acquisition means that, in theory, more data can be recovered. At a high level, therefore, time to acquire $Clock_{27}$ and the number of packets captured thereafter were hypothesised as useful metrics to gauge success.

To assist in assessing congestion of the ISM band, a suitable model was found in the paper by \cite{Gummadi2007}, which carried out practical experiments examining RF interference on Wi-Fi networks from ISM band sources. This paper is not included in the literature review, as it does not offer any particular contribution to developing capture of Bluetooth, however, the experimental mechanism appears to be adaptable to the analysis of RF congestion in a Bluetooth setting, and the subsequent effects on capture rate.

In common with BlueEar \cite{albazrqaoe2016practical}, the authors use a two radio setup – one to perform the specific experimental activities, in this case, monitoring the Wi-Fi throughput in response to Zigbee and other ISM traffic, and the other to measure the level of RF interference.

As this method aligns with that chosen by other Bluetooth researchers, a two-radio method was chosen for the experimental activities of this project. \cite{Gummadi2007} provide a second useful pointer – in the description of their experimental setup, they describe carrying out 10 experimental runs for each test. This model was adopted for experimentation, to provide a meaningful amount of data for analysis. 

Various mechanisms were discovered through experimentation to assist in making the experiments repeatable, with the intent of eliminating potential factors which could skew the results, or make it hard to compare the results from one run to another. 

The final capture process involved fully resetting the environment before each run, which involved:
\begin{itemize}
\item Switching the smartphone devices into flight safe mode to clear all connections;
\item Unplugging the Ubertooth devices from USB;
\item Switching power off using the car’s ignition key – it was determined that the AV system powers off after 30 seconds in this state when not being used for radio/media playback; and
\item Power off speaker systems (such as the Bose Soundlink).
\end{itemize}

The steps were repeated in reverse to ready the environment for the next capture run. This process was mildly cumbersome, but was performed to ensure that information retained by the Ubertooth device from a previous run was not able to influence the next. Previous researchers in the field have not described any steps taken to isolate experiments from each other in this fashion – this may indicate that such steps are not required, or may indicate that the authors did not feel this information added anything to their published results.

\subsection{Experimental Setup - Overview}
\begin{figure}
  \includegraphics[width=\linewidth]{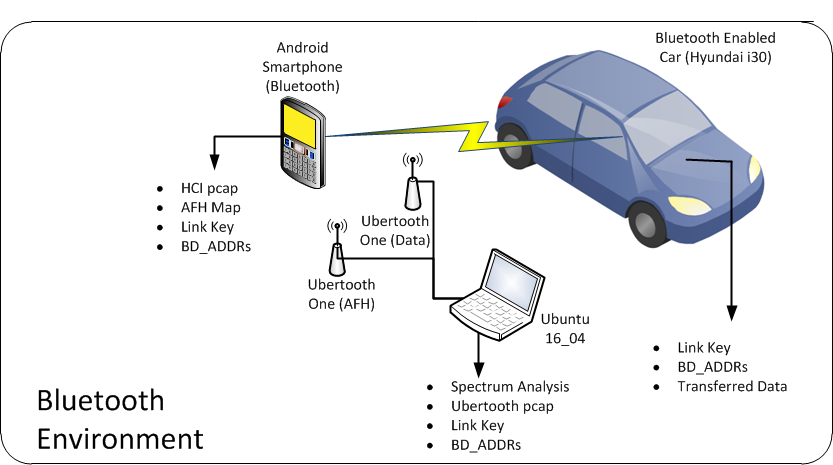}
  \caption{Experimental Capture Environment – vehicle test}
  \label{fig19}
\end{figure}

Two experimental setups are proposed. For testing against a vehicle, the environment will be configured as shown in  Figure \ref{fig19}, with a smartphone used to initiate Bluetooth connections to the vehicle, whilst being monitored by a laptop with two Ubertooth One devices. 

Using a similar “two radio” setup to \cite{albazrqaoe2016practical}, one Ubertooth device was used to monitor the RF environment, capturing the AFH map of the test piconet. In line with the method described by \cite{Gummadi2007}, the AFH Map is captured once per second. The second Ubertooth device was used to passively snoop the audio data being transferred.

\begin{comment}
Using a laptop to generate the traffic was considered, and may have provided more flexibility. 

In practice, the limitations of available laptop hardware and logistics of operating multiple items of equipment in a car forced the decision to use a smartphone as the Bluetooth traffic generation device. 

An unfortunate discovery followed - the Bluetooth stack used on Android smartphones was changed from the Linux derived BlueZ stack to the Broadcom authored BlueDroid from Android 4.2 onwards \cite{Holtmann2014}. BlueZ is more flexible, and provides useful tools; in particular hci-tool and hci-config which can report the AFH map currently in effect, and enable or disable AFH behaviour respectively. This change in stack removed the ability to capture the AFH map from the smartphone device, which proved to be a significant obstacle to verifying some of the experimental results. 
\end{comment}

\begin{figure}
  \includegraphics[width=\linewidth]{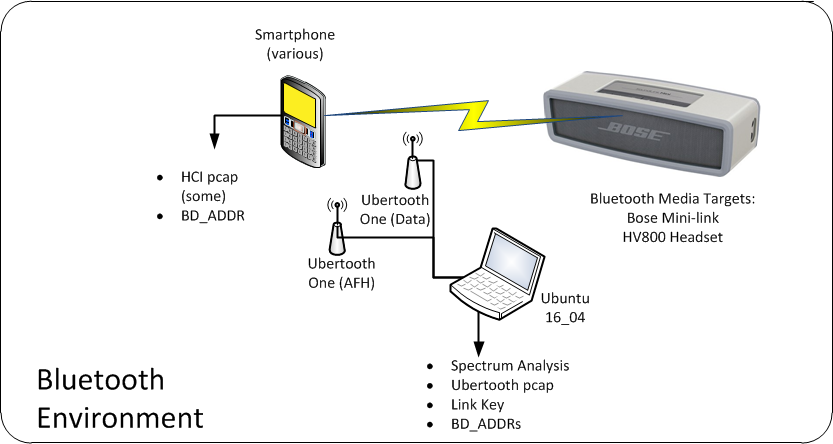}
  \caption{Lab Based Experimental Setup}
  \label{fig20}
\end{figure}

To produce a consistent, repeatable, stream of data the smartphone media player was used to play the same audio track, starting fresh for each test run. The author’s sanity was not being tested as a factor in this experiment. A second, lab based environment was used to perform tests against commonly available consumer Bluetooth media devices – a speaker, and a headset. This is shown in Figure \ref{fig20}.

This second environment was used for two distinct purposes – firstly to provide a more consistent “noisy” RF environment for tests where manipulation of the AFH map was attempted, and secondly to determine to what extent the use of devices with older and newer Bluetooth versions affected the ability to capture data, where RF noise was not an experimental factor being measured. 

\subsection{Devices Used}
\begin{figure}
  \includegraphics[width=\linewidth]{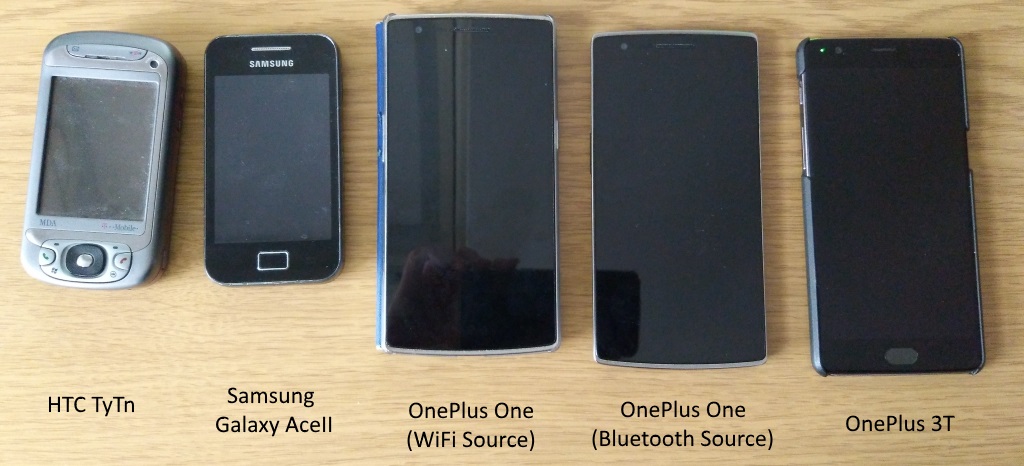}
  \caption{Smartphones used for Experimentation}
  \label{fig21}
\end{figure}

Five smartphones were used in total (Figure \ref{fig21}), with the configuration and characteristics summarised in Table \ref{tab:4}. The oldest was a 2006 HTC “TyTn”, running Windows Mobile 5, and supporting Bluetooth 2.0. This phone was of limited value, as it provided very little in the way of accessible tools or diagnostic information; however, it was used in a single experiment (Experiment 2) to determine the relative susceptibility of older Bluetooth 2.x implementations to snooping, relative to the newer 4.x versions. The other smartphones used were all Android devices, the oldest of which was a 2011 Samsung “Galaxy Ace”. This phone was a Bluetooth 3.0 chipset, and originally used Android 2.3.

\begin{table*}[t]
  \caption{Bluetooth Devices Used in Experiments}
 \begin{tabular}{p{1cm} | p{1cm} | p{1cm} | p{3cm}  | p{3cm}  | p{3cm}  | p{3cm}}
Device	& LAP	& UAP	& OS	& OSVer	& Version & Notes \\
\hline 
OP3T	& fd7fd1	& fb	& Android	& 7.1.1	& 4.2	& 	OnePlus 3T \\
ACEII	& 214b5f	& a4	& Android	& 4.4.4	& 3.0	& 	Samsung Galaxy ACE II \\
WM5	& 392795	& 76	& Win Mobile	& 5.1.195	& 2.0	& 	HTC TyTn \\
Bose	& 24cb9d	& 1f	& N/A	& N/A	& 2.1	& Bose SoundLink Mini \\
i30	& 198626	& 44	& Win CE	& 6.0 CE	& 2.1	& 2013 Hyundai i30  \\
OPO	& 3456fa	& fb	& Android	& 6.0.1	& 4.1	OnePlus One \\
APTx	& 600df9	& db	& N/A	& N/A	& 2.1	& HV-800 Stereo Headset \\
\end{tabular}
  \label{tab:4}
\end{table*} 

To provide the means to capture HCI traffic, this phone was ‘rooted’, and flashed with a newer Android version, 4.4.4, using the CyanogenMod project’s CM11 build. The phone was more useful than the TyTn, but proved to be limited due to the Broadcom chipset used in the device; this implementation required a closed source binary driver which limited the ability of the hci-tools suite to provide useful information. 

A pair of newer smartphones from the manufacturer OnePlus were also used – a 2014 OnePlus One running Android 6 and supporting the Bluetooth 4.1 standard, and the newest phone used, a 2016 OnePlus 3T, running Android 7 and supporting Bluetooth 4.2. Each of these phones was used to gauge relative performance between older and newer Bluetooth implementations, but were also used to measure the capture rates in noisy and quiet RF environments. 

To evaluate whether the simple secure pairing (SSP) modes introduced in Bluetooth 1.2 made a measurable difference in capture rate, two different media targets were used. One, a Bose Mini Soundlink, supports Bluetooth 2.1 configured with a default passphrase of ‘0000’, and the other, a cheap Bluetooth headset based on the Cambridge Silicon Radio (CSR) chipset is more basic – similarly a Bluetooth 2.1 device, this supports SSP in the “just works” configuration – in theory the weakest and simplest pairing schema available.  To remove other potential factors from this experiment (Experiment 3), only a single handset was used – the OnePlus One, and the experiment was only performed in an RF Busy environment. 

\subsection{Scenarios Tested}
\begin{figure}
  \includegraphics[width=\linewidth]{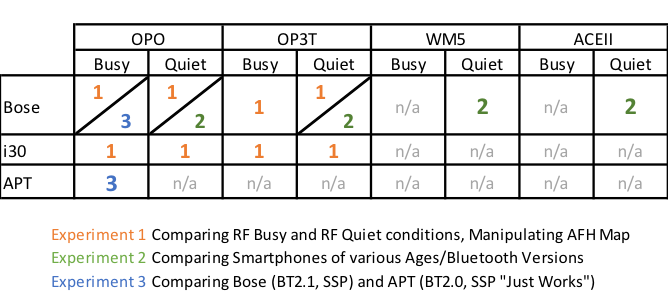}
  \caption{Devices and Scenarios tested in each Experiment}
  \label{fig22}
\end{figure}

Three experimental scenarios were settled on, as shown in Figure \ref{fig22} - for each experimental run, the following information was recorded:

\begin{itemize}
\item The start time of the run;
\item How long it takes to find $Clock_{27}$ (if successful);
\item How many guesses were required to determine the clock;
\item How long it takes to successfully decode a packet (again, if successful); and
\item Statistical information about the capture: Number of decoded packets; Number of ‘NULL’ or ‘POLL’ packets; and Number of failed decode attempts, after the clock is established.
\end{itemize}

\subsection{Wi-Fi Environment – From Quiet to Reliable Congestion}
For each of the Bluetooth experimental environments, a common Wi-Fi test environment was used. 

\begin{comment}
 \begin{figure}
  \includegraphics[width=\linewidth]{figures/image023.png}
  \caption{Wi-Fi Congestion Environment (all experiments)}
  \label{fig23}
 \end{figure}  
\end{comment} 

A second smartphone, connected to a Wi-Fi access point, was used to stream media from a local server to create a predictable, repeatable level of RF congestion, and hopefully to induce consistent AFH behaviour.
It would have been possible to use the same smartphone as was being used to generate Bluetooth traffic, however, as described by \cite{Chokshi2010}, the Bluetooth and Wi-Fi functionality in devices which support both technologies is most typically provided by a single Baseband System on a Chip (SOC). Using separate devices ensures that any potential interference is due to effects within the RF environment, similar to those an attacker would experience, rather than due to resource congestion or shared access to the radio device within a single smartphone. In every case, a second 2014 OnePlus One handset running Android 6 was used to generate the Wi-Fi traffic. It should be noted – this was not the same device used to generate Bluetooth traffic.
In configuring and validating this environment, a single Ubertooth was used, along with the Kismet Spectrum Analyser Tools (spectools), this uses the Software Defined Radio (SDR) to generate a real time stream of signal strength information across the ISM band. The output of this device is shown in Figure 24 below . In this output, the channel numbers displayed below the Spectral View represent the midpoint of each Wi-Fi channel. The main channel in use is channel 6, and there is some minor traffic on channel 13. 

\begin{comment}
\begin{figure}
  \includegraphics[width=\linewidth]{figures/image004.png}
  \caption{Wide 40MHz Wi-Fi Channel Usage with Bluetooth Audio Playing}
  \label{fig24}
\end{figure}

Figure \ref{fig24} also demonstrates an issue found in early testing. Wi-Fi access points compliant with 802.11n can use either a 20MHz channel or 40MHz channel – in effect, up to 2/3 of the ISM band. As this broader channel usage can be significantly more prone to interference, Wi-Fi access points configured for a 40MHz channel width will back off more readily than those configured for 20MHz to avoid unwanted interference with other devices. 

In practice, it proved difficult to reliably saturate a broader 40MHz channel with Wi-Fi traffic – as visible in Figure \ref{fig24}. To provide a more consistent Wi-Fi environment, the channel width on the test Wi-Fi access point was forced to 20MHz (Figure \ref{fig25}). 

 \begin{figure}
  \includegraphics[width=\linewidth]{figures/image024.png}
  \caption{Wi-Fi Access Point Channel Width Selection}
  \label{fig25}
\end{figure}   

Once this was carried out, the
\end{comment}
Wi-Fi throughput could be sustained at a consistent level of around 40Mbps. This was achieved by copying a large ISO file from the server to the smartphone. The same file was copied during all tests, and the copy was restarted each time. This ensured that similar traffic was being generated during “RF Busy” tests, although, it would emerge that the behaviour of AFH and the interaction between Bluetooth and Wi-Fi did not allow for this high throughput or, indeed, consistency from one run to the next.

Producing a quiet RF environment for testing was challenging, but possible. As the key experimental activity is to evaluate the effect of a congested RF space map, it was important to be able to establish a baseline of performance in a repeatable, measurably quiet environment. Early experimentation highlighted that both other vehicles passing and the vehicle’s own non-Bluetooth systems generated a sufficient level of radio interference on the ISM band to influence the AFH data being captured.

\begin{figure}
  \includegraphics[width=\linewidth]{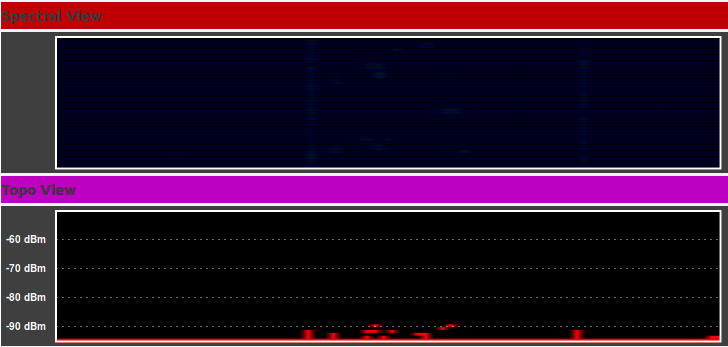}
  \caption{RF environment used for "Quiet" tests - no Wi-Fi or external Bluetooth activity}
  \label{fig26}
\end{figure}

To avoid these potential sources of interference, RF Quiet tests in a vehicle setting were therefore carried out with the vehicle parked far away from buildings, with the engine off. All Wi-Fi devices were disabled, and potential Bluetooth sources (such as the author’s personal fitness tracker) were removed from the environment, producing the RF Quiet environment seen in  Figure \ref{fig26}. 
 
\begin{figure}
  \includegraphics[width=\linewidth]{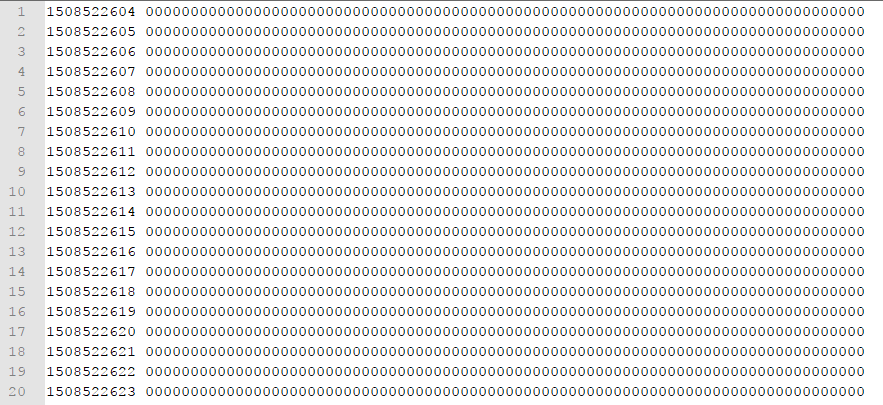}
  \caption{AFH Map output from Ubertooth Tools during preparation for "RF Quiet" Tests}
  \label{fig27}
\end{figure}

The RF environment was inspected prior and subsequent to each test and verified by examining the AFH\_Map  (Figure \ref{fig27}) to ensure that the quiet conditions were maintained. 
 
The requirement to have the vehicle’s engine turned off was particularly troublesome, as this limited the amount of time available for testing before the engine had to be restarted to prevent the battery from becoming too deeply discharged.

\subsection{Capture and Analysis Tools}
\begin{figure}
  \includegraphics[width=\linewidth]{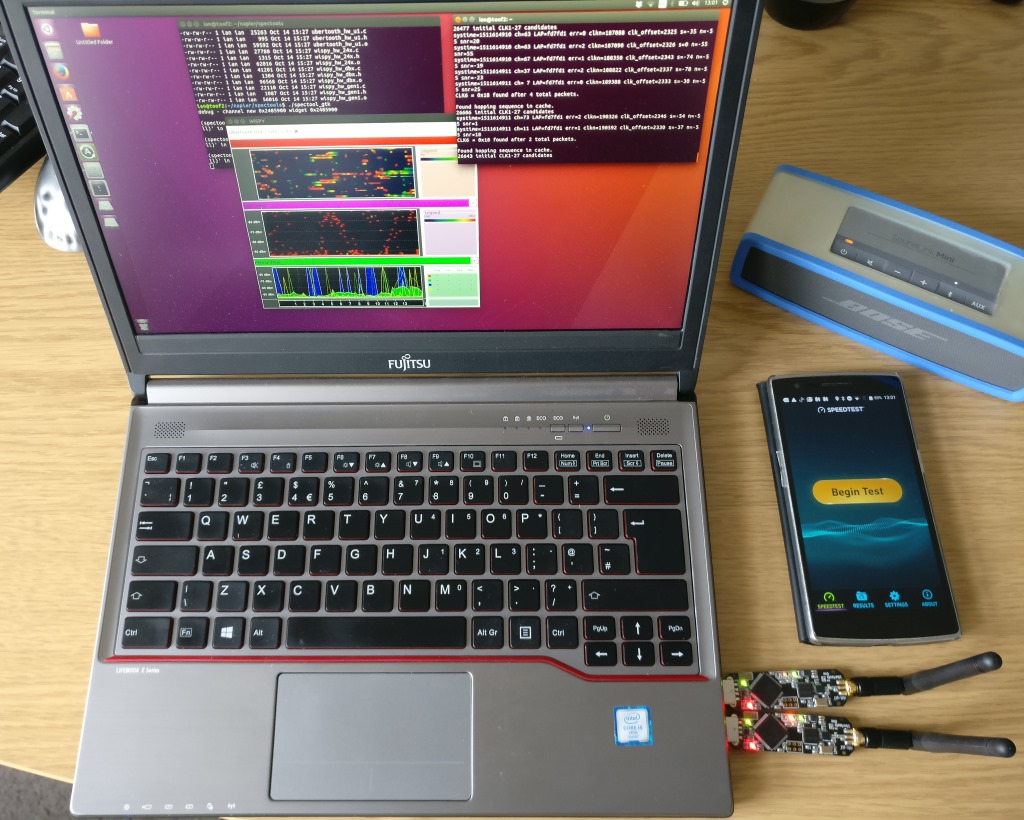}
  \caption{Desk Based Lab Setup}
  \label{fig30}
\end{figure}

As discussed above, capture activities were performed using a pair of Ubertooth One devices, following the two radio approach of \cite{albazrqaoe2016practical} and \cite{Gummadi2007}. The Ubertooth tools used were based on version 2017-03-R2 pulled from GitHub and compiled on the test laptop (Figure \ref{fig30}), a generic x64 Intel machine running Ubuntu Xenial 16.04. Initial tests were performed using the more security focused Kali 2016\_r2 \cite{Muniz2013}, however, the rolling updates of this system did not provide a stable enough build environment for the Ubertooth tools, which have a dependency on older libusb versions.

The 2017-03-R2 version of the tools incorporates the initial BlueSniff code, with improvements described in \cite{Spill2012} to incorporate AFH and following behaviour, and further improvements to the codebase around clock detection that were proposed in BlueID. As the capture of data is dependent on acquiring $Clock_{27}$, the improvements in clock detection make the overall capture rate more successful \cite{Huang2014}.

Wahhab Albazrqaoe, one of the authors of the Michigan State University papers, was contacted and kindly provided the Source code for the more advanced version of the Ubertooth tools described in \cite{albazrqaoe2016practical}. It was hoped that this would allow for the capture mechanism described in BlueEar to be repeated, however, on surveying the supplied code, it became apparent that the BlueEar code is based on the Ubertooth project’s earlier 2015-10-R1 release, and was therefore not directly compatible with the tools being used for experimental capture. 

\begin{figure}
  \includegraphics[width=\linewidth]{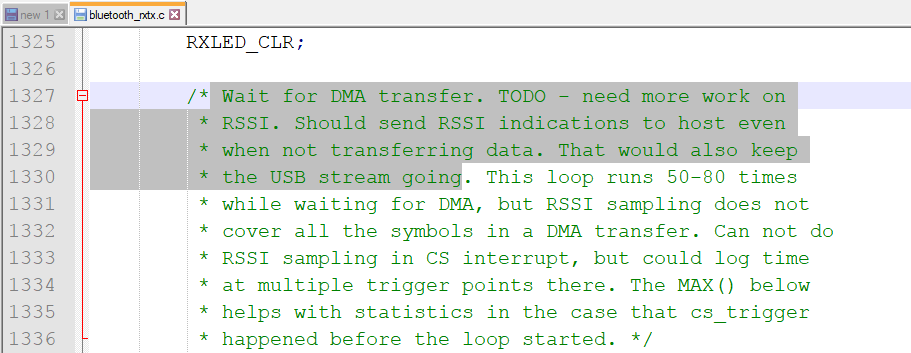}
  \caption{BlueEar Code, indicating de-scoping of RSSI data stream from code.}
  \label{fig31}
\end{figure}

BlueEar works by replacing the firmware code in bluetooth\_rxtx.c with code which uses the techniques described in BlueEar to more accurately model the remote AFH map. This updated firmware allows one Ubertooth to be designated as the “Scout”, which provides an accurate model of RF channel usage, and therefore an accurate estimation of the AFH Map moment by moment. The other Ubertooth is designated as the “Sniffer”, and performs the actual capture activities. The devices are dedicated to this functionality and as such, some of the original Ubertooth functionality is lost. Of concern for this project, the ability to provide real-time RSSI data appears to have been removed, as highlighted in Figure \ref{fig31}.

For the planned experimental activities, the ability to use the spectools spectrum analyser and Kismet packet capture environment in addition to ubertooth-rx was required. It was therefore decided to forgo the potential improvements in capture rate offered by BlueEar, to maintain the flexible range of tools available for use. 

Prior to performing the experiments, the BD\_ADDR of each participating device was discovered and recorded. This allowed the LAP and UAP for each piconet to be identified in advance, and removed the need to run a ‘survey’ activity to identify the UAP for each experimental run. 
To capture Bluetooth traffic, and an associated AFH map, an approach similar to that used by \cite{albazrqaoe2016practical} is deployed – two Ubertooth devices are used; one attempts to capture the AFH map of the piconet using the ubertooth-afh tool, whilst the other performs a sequence of data capture activities using the ubertooth-rx tool. 
ubertooth-rx was used in a time-bounded mode where it runs for a period of 180 seconds, and attempts to:

\begin{itemize}
\item Use the provided LAP and UAP to determine which Piconet is being monitored;
\item Perform the brute forcing of $Clock_{6}$ described in BlueSniff;
\item Once $Clock_{6}$ is discovered, generate the entire hopping sequence;
\item Test possible $Clock_{27}$ candidates, using the brute force approach of \cite{Spill2012}; and
\item Once (if) $Clock_{27}$ is acquired, follow the piconet, and attempt to capture subsequent packets.
\end{itemize}

The ubertooth-afh tool was used to capture the AFH map of the piconet once per second as per \cite{Gummadi2007}. 
Dominic Spill, author of BlueSniff and lead developer of the Ubertooth project, was contacted and provided some useful pointers in how to modify the Ubertooth code. The underlying libubertooth was modified by the author to include a system timestamp to allow the AFH map to be compared to the ubertooth-rx output. This involved a relatively simple change to the ubertooth\_callback.c component of libubertooth. 

 \begin{figure}
  \includegraphics[width=\linewidth]{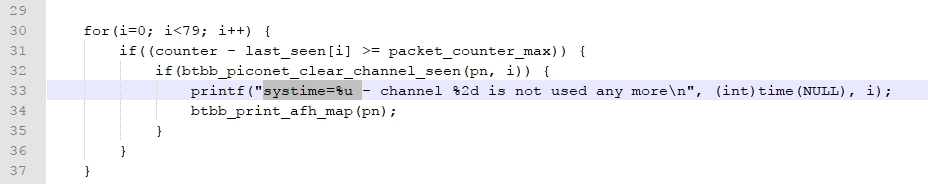}
  \caption{Adding timestamp outputs to the Ubertooth AFH tool}
  \label{fig32}
\end{figure}  

Once altered, the make environment was reset, and the tools rebuilt from source. All experiments were performed using this modified version of the ubertooth-afh tool. 

\subsection{Tools Used in each Experiment}
Each Ubertooth tool was run in a separate terminal window, displaying the output to the console, and simultaneously capturing to a text file using tee:

ubertooth-rx -l fd7fd1 -u fb -U 0 -t 180 | tee run1.console
ubertooth-afh -l fd7fd1 -u fb -U 1 -r | tee run1.afhmap

In this example, the piconet in which the master device has LAP fd7fd1 and UAP fb is monitored (this is the OnePlus 3T smartphone). 
The parameters are as follows:
\begin{itemize}
\item -U directs each tool to use a separate Ubertooth device;
\item –t 180 parameter causes the capture to terminate after 180 seconds; and
\item –r parameter tells ubertooth-afh to export the currently observed AFH Map once per second, in the binary format shown in (Figure \ref{fig32}) to a file called runx.afhmap. 
\end{itemize}

In this representation, Channel 0 is output after the timestamp, with one digit representing each Channel – 0 means the channel is ‘unknown’ and available for Bluetooth to use, 1 means the channel is ‘bad’ and removed from the Hopping Set.

 \begin{figure}
  \includegraphics[width=\linewidth]{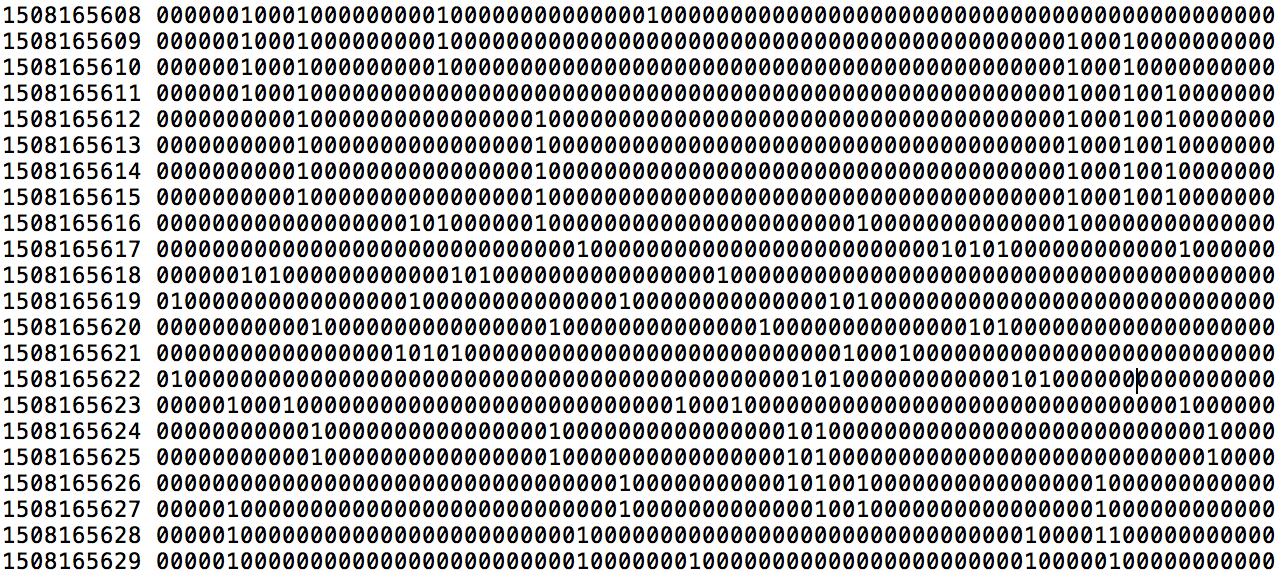}
  \caption{Binary AFH Map Capture format}
  \label{fig33}
\end{figure}   

The map shown in Figure \ref{fig33} was captured during run 1 of a capture session using the OnePlus 3T handset and Hyundai i30 in an RF Quiet setting, and is broadly typical of those captured during these experiments. 

\begin{figure*}
  \includegraphics[width=\linewidth]{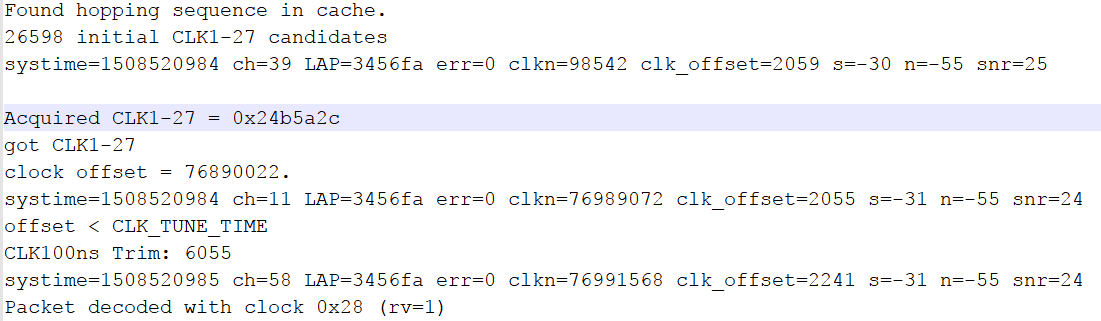}
  \caption{Full $Clock_{27}$ Acquisition during a successful capture run}
  \label{fig34}
\end{figure*}

Alongside the AFH Map, the console output of ubertooth-rx was captured to a file called runx.console. This console file was then parsed to find the timestamp for events of interest, notably the discovery of the full clock, $Clock_{27}$ which allows for the entire hopping sequence to be calculated. This event, as an example of runx.console output is shown in Figure \ref{fig34}.

For each experimental scenario, the capture session was reset and run 10 times, as per \cite{Gummadi2007} with all log files being retained for analysis. 

\subsection{Metrics of Capture Quality and Success}
Various metrics were identified which could be extracted from the console log files that could be used to weight the relative success of one capture attempt to another. The first of these metrics is the number of guesses required to discover the full clock. This was chosen as a measure, reflecting the work in \cite{Huang2014}, where timing information of packets received was used to fingerprint devices, even when the packets themselves could not be decoded. 

Once the Ubertooth tool successfully extracts $Clock_{6}$ from a packet, it calculates the entire hopping sequence, and begins to pattern match the hops of incoming packets to determine the current offset.  This event is indicated in the console logs by the text “x initial CLK1-27 candidates”, where x is a number, typically around 26,400. If the tool encounters enough incorrect matches to conclude that none of the proposed $Clock_{27}$ candidates was correct, the guessing process is reset and a new candidate value sought. 

If the clock is correctly guessed, this provides the second metric; the time taken to discover the full clock, which was calculated by the difference between the first “systime” timestamp in the capture file, and the timestamp of the packet where the text “Acquired CLK1-27” appears. If this text was not present, then the capture was considered a Fail and no packets can be decoded.

The time taken after acquiring $Clock_{27}$ till decoding the first data packet was also recovered from the timestamps, however, in practice this was always within a second or so.

\section{Results and Evaluation}
\subsection{Introduction}
Having considered mechanisms which could test the hypothesis around AFH, and settled on the experimental setup detailed above, the experiments were run to produce data as consistently as possible.

For each of the capture runs, the timing metrics identified previously were extracted. In addition, statistical information to assist in understanding the quality of data capture were gathered from the log files, including:

\begin{itemize}
\item The number of successful packet decodes.
\item The number of failure decodes.
\item The number of good data packets decoded (as opposed to NULL/POLL packets). 
\end{itemize}

Once gathered, the data was examined to determine to what extent a Busy RF environment impacted on data capture rates, and whether the hypothesised approach of \cite{Spill2007} and \cite{albazrqaoe2016practical} was able to provide a measurable improvement.

\subsection{Experimental Data}
An example of the data gathered is shown in raw format in Figure \ref{fig35}. The experimental data gathered is included in full in Appendix 1. Ultimately, the capture runs exhibit very large time differences, from a minimum of four seconds through to a maximum of 2 minutes 55 seconds; with the likelihood that several of the captures which terminated at the 180 second/3 minute mark may well have succeeded if allowed to run beyond this time. 

 \begin{figure*}
  \includegraphics[width=\linewidth]{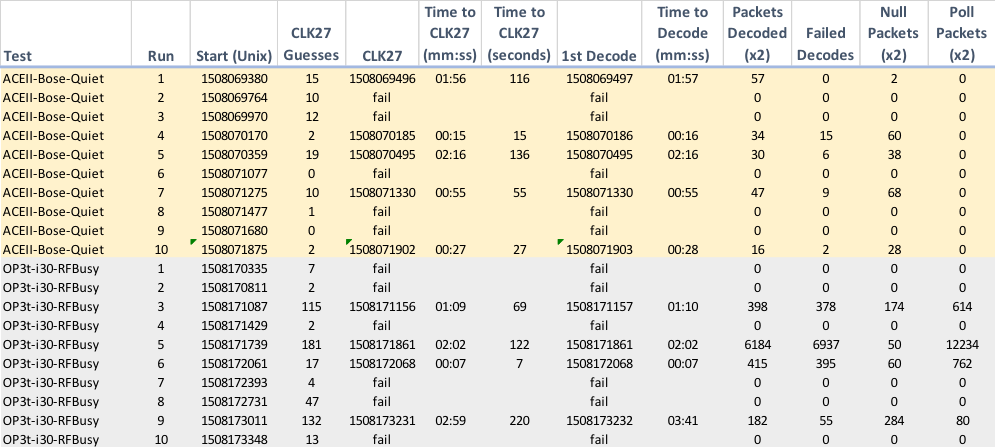}
  \caption{Raw Data gathered from console dump files}
  \label{fig35}
\end{figure*}    

\begin{figure}
  \includegraphics[width=\linewidth]{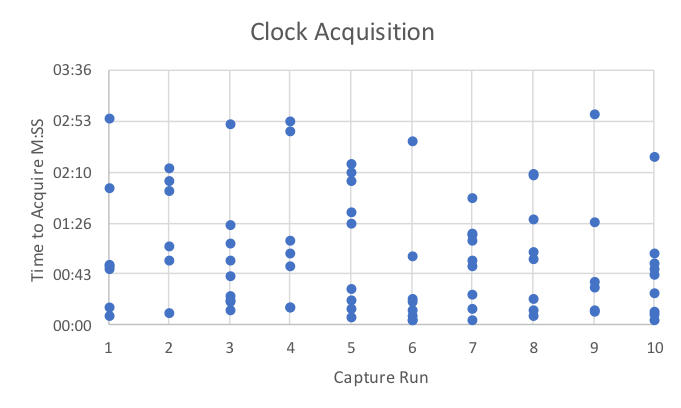}
  \caption{Time for complete clock acquisition per Run (All Experiments)}
  \label{fig36}
 \end{figure}

Figure \ref{fig36} charts the time in minutes and seconds to acquire the full clock. Each data point represents a single capture run, and as this graph is intended to simply demonstrate the variation in acquisition time, all of the experimental scenarios are overlaid on the same graph, such that each scenario’s “Run 1”, “Run 2” and so on are grouped. 
 
Given the variability of clock acquisition time, it is entirely possible that this does not represent a good proxy for the success of one capture run over another. In any case, as the divergence between runs is evident at the scale of seconds, adding millisecond resolution would not appear to add any additional clarity to the results.

\begin{figure*}
  \includegraphics[width=\linewidth]{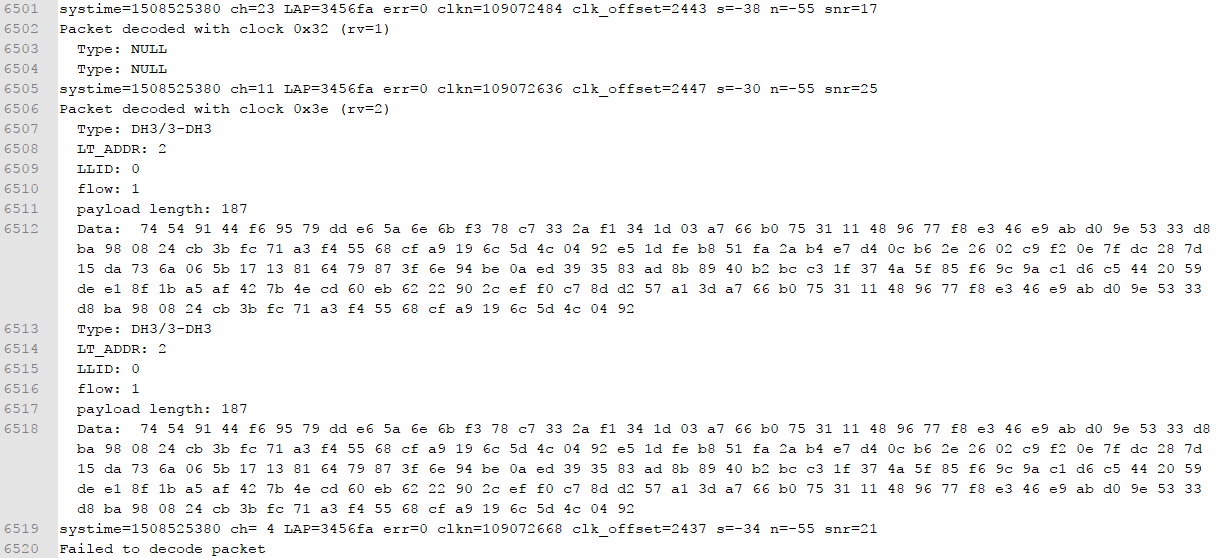}
  \caption{Three outcomes of a captured packet: decoded, NULL, or Failed}
  \label{fig37}
 \end{figure*}

In addition to timings, the other information drawn from the console logs is a measure of the quantity of data recovered, and an indication of data quality. Once the full clock has been acquired, and both the hopping sequence and current offset determined, the Ubertooth tools are able to hop along with the Bluetooth piconet, and recover data packets. Each time a captured packet is analysed, a console entry is created, with the outcome and, if successful, an extract of the data from the packet, shown in Figure \ref{fig37}.

Figure \ref{fig37} shows the three potential outcomes. The packet captured at line 6501 is successfully decoded with a matching clock offset, however the packet is a NULL type – one of two heartbeat style packets (the other being POLL) which contain no useful data. Bluetooth devices send a packet on their allocated slot, whether they have meaningful data to send or not, and this results in a large percentage of the received packets comprising of these POLL/NULL packets. In the experimental results of this project, the percentage of NULL/POLL packets ranged from 10.36\% to as high as 96.04\%. 

The second packet, captured on line 6505, is also decoded, and contains valid data – a 3-DH3 packet. This is a three slot long packet which is part of the inquiry/response mechanism used to relay device capabilities \cite{BluetoothSIG2007} . In the playback of audio, these appear to be used to support volume control adjustments between the devices. In the graphs and later descriptions, packets which were able to be decoded in this fashion and contained valid data  are collectively described as “good data”.

The third outcome is a failed decode – the packet captured on line 6519 is recognised as part of the piconet, and has an appropriate sequence, however, the data itself was not able to be recovered. 

\begin{figure*}
  \includegraphics[width=\linewidth]{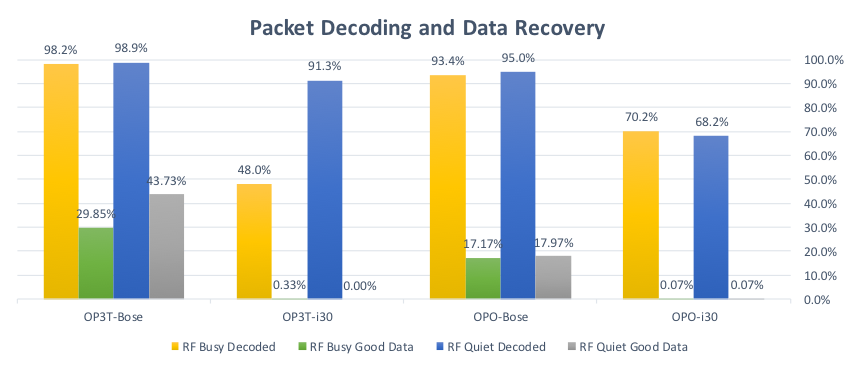}
  \caption{Averaged Rates of Packet Decoding and Data Recovery (Experiment 1)}
  \label{fig38}
 \end{figure*}

Figure \ref{fig38} represents the averaged rates for each outcome across the capture combinations involving the OnePlus One and OnePlus 3T handsets, connecting to the Bose Soundlink and Hyundai i30, with each combination being tested in both RF Busy and RF Quiet scenarios (Experiment 1).

It appears that achieving a high percentage of received packets being decoded does not necessarily correlate with a larger percentage of good data being recovered. In each case where the Hyundai i30 was the target device, a relatively successful rate of packet decoding nonetheless resulted in almost no data being recovered. 
This analysis provides the first finding of this project; Only a small percentage of transmitted data seems to be recoverable from radio signals, and this points towards a dependency on the transmitting device using Basic Rate data transmission. 

\subsection{Examining Data Capture Rates}
\begin{figure*}
  \includegraphics[width=\linewidth]{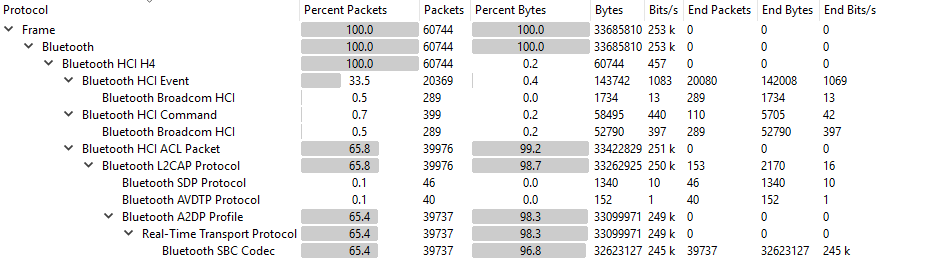}
  \caption{Wireshark Analysis of captured HCI Dump}
  \label{fig39}
 \end{figure*}

Comparison with the HCI-dump pcap file generated on the OnePlus handsets demonstrates how large the shortfall is between the data transmitted and data captured. As a representative example, the Pcap file generated by a test run of the OnePlus 3T handset against the Bose Soundlink of captured data from Ubertooth was around 30-40Kb in size, with 300 packets captured.  The corresponding HCI dump from the OnePlus 3T itself was 34.3Mb in size, and contained 67,000 packets (Figure \ref{fig39}). 

Inspection of these recovered packets, and comparison with the captured packets reveals another potential issue for would-be packet sniffers, and highlights a possible weakness in the experimental approach; it appears that the packet types most likely to be successfully captured from the air are HCI Event packets, specifically DH3 and DH5 packets. Notably, these packets are modulated using the simpler GFSK modulation that is used by Bluetooth Basic Rate, rather than the considerably harder to demodulate QAMFSK used in Bluetooth Enhanced Data Rate (EDR). 

A large (greater than 65\%) proportion of the traffic is comprised of HCI ACL Packets, only a very small number of which will be accessible to the Ubertooth to capture. As described previously, the evolution of the Bluetooth standard to support higher data rates involves the use of more complex modulation schemes. Capture of a frequency hopping signal requires an assessment of whether a detected radio signal in a given channel slot represents a meaningful signal to be demodulated and decoded, or random radio noise. Where the communication is between two parties who have pre-negotiated the hopping sequence, the hopping sequence is known; therefore, the receiver has a higher degree of confidence that a guess that channel x contains data rather than noise is likely correct. 

The third party observer, on the other hand, does not have the sequence to start with, and must therefore make guesses in a more complex and error prone environment. Audio playback was used as a means to generate a steady, consistent stream of data for long enough to be captured. The use of stereo audio, however, may have produced a data rate high enough to require the Enhanced Data Rate (EDR), a feature since Bluetooth 2.0, meaning that the traffic able to be captured and analysed was significantly (and unintentionally) reduced \cite{Naggs2013}.

It is notable that \cite{Spill2007}, \cite{Huang2014} and \cite{albazrqaoe2016practical} all describe the capture environments of their experimental setups in detail, but do not explain how they generate traffic to be detected and sniffed. This lack of detail makes comparison with the method used in this project difficult. EDR is not mentioned in Spill’s 2007 paper, however, he does mention that it should be possible to capture in his Usenix WooT presentation that year. It appears, in light of the results obtained in this project, that this was an aspirational goal of the Ubertooth project, rather than an in-development capability. The Ubertooth toolset represents the best efforts of researchers so far, however, even in the 2017-03-R2 release of Ubertooth tools, the authors note that Bluetooth Classic Basic Rate capture is supported, whilst Enhanced Data Rate (EDR) capture is ‘experimental’ at best. The experimental results seem to show that the data which is being generated in order to be sniffed is an important factor – to produce useful data, a transmitting device must only be generating traffic using Basic Rate signalling. As this is highly unlikely to be the case for an arbitrary device which an attacker wishes to intercept, this in itself is likely to indicate that passive snooping presents less of a risk than some researchers have previously indicated. 

In any case, whilst the quantities of data being captured are low, and represent the control channel of communication, rather than the actual data itself, this capture happens at a high enough level to provide comparison between different devices and situations. 

To test the assumption that newer Bluetooth protocol enhancements have made capture more difficult, the results were examined to determine to what extent there was a measurable difference in the capture characteristics of older devices, and whether these were easier to reliably capture than the newer devices. 

 \begin{figure}
  \includegraphics[width=\linewidth]{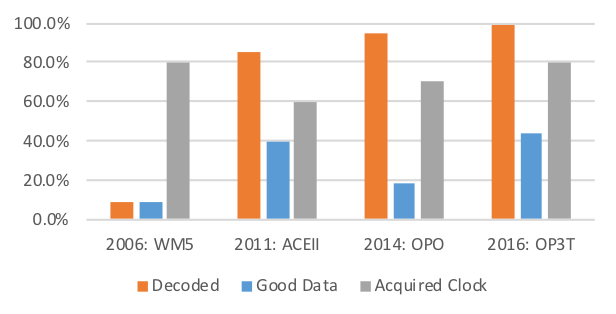}
  \caption{Comparative Data Capture and clock acquisition by Device Age (Experiment 2)}
  \label{fig40}
 \end{figure}   

In contrast to the expected behaviour, testing showed that the oldest devices were not demonstrably easier to capture data from – on none of the metrics is there a significant correlation between the age of the device and an improved rate of capture (Figure \ref{fig40}).  As it appears that the use of Enhanced Data Rate traffic may represent a key difficulty in capturing data, and this has been part of the standard since Bluetooth 2.0, only a very narrow time window of devices, supporting Bluetooth 1.0, 1.1 or 1.2 would in theory be more susceptible.

\begin{comment}
What does appear to be the case, however, is that certain pairings are more amenable to capture than others. As Figure \ref{fig41} shows, the OnePlus 3T handset produced particularly poor results when paired with the Hyundai i30 – second only to the Windows Mobile 5 device. In Contrast, the OnePlus One Handset gave some of the most reliable captures of the experimental run in terms of clock acquisition. 

 \begin{figure}
  \includegraphics[width=\linewidth]{figures/image041.png}
  \caption{Good Clock Acquisitions (out of 10 Experimental Runs)}
  \label{fig41}
 \end{figure}   
\end{comment}

Busy RF conditions do appear to be slightly advantageous for more rapidly guessing $Clock_{27}$, however, this does not seem to confer any benefit in terms of actual data capture.

\subsection{Excluding Bluetooth from a given Spectrum with Wi-Fi}
While it was not possible to recover substantial quantities of data for packet analysis, the Ubertooth devices proved to be useful for capturing spectrum usage information. Having discovered that specific pairings of device behaved consistently, two of the smartphones and two of the targets were analysed together, attempting to repeat the capture experiment using each in RF Quiet, and RF Busy conditions (Experiment 1).

In BlueEar, \cite{albazrqaoe2016practical} attempt to use a better knowledge of the AFH environment, to improve capture rates – they are able to show a better rate of capture in busy environments. Whilst not using the BlueEar code, for the reasons described above, this experiment seeks to verify the same behaviour, using Wi-Fi.
The Wi-Fi experimental apparatus was used to generate a steady stream of traffic by copying ISO images from a network share to the smartphone. After an initial surge of speed, as the server’s cache was exhausted, this settled to a steady 41MBps, as measured by the smartphone. 

 \begin{figure}
  \includegraphics[width=\linewidth]{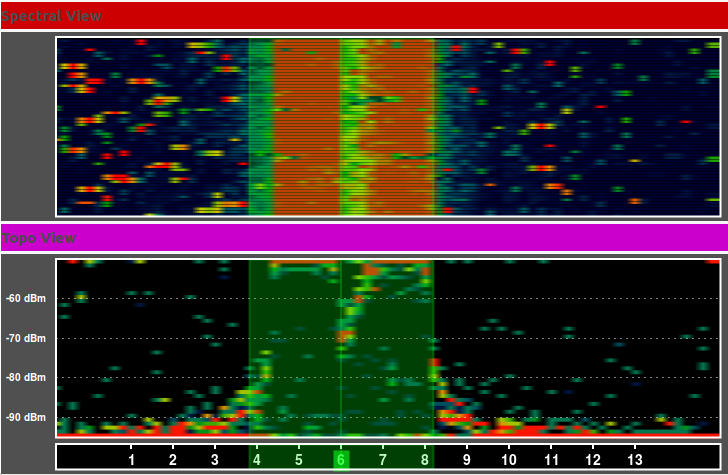}
  \caption{Wi-Fi channel 6 saturated, then Bluetooth playback started}
  \label{fig42}
 \end{figure}   

Once the Wi-Fi throughput was stable, Bluetooth audio playback was started with the pairing being tested. Kismet spectools was used to sample the RF environment, producing the output shown in Figure \ref{fig42}. This includes the previously described spectral view, and one other – the “Topo” view. In this view, the X-axis again represents all channels from 0 at the origin to 79 at the far right, and the Y-axis represents the sum of signals observed. Over time, in the absence of new signals, the pixels plotted will gradually fade to black, and move down until they drop below the -90dBm “floor”. As new signals are observed in the same channel, however, they grow in intensity from ‘cold’ green to ‘hot’ red, and are plotted at their observed signal strength. 

in Figure \ref{fig42}, we see almost complete saturation of the 20MHz of spectrum around the centre of Wi-Fi channel 6, with occasional ‘flecks’ elsewhere – these single pixel green RSSIs are a visual representation of the Bluetooth audio traffic. During this period, Wi-Fi performance dipped slightly when Bluetooth playback was started, dropping to a sustained 38MBps. 
 
 \begin{figure}
  \includegraphics[width=\linewidth]{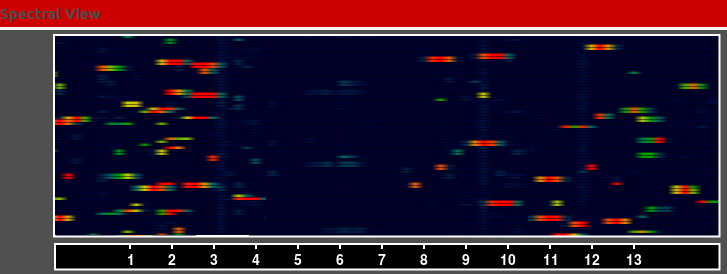}
  \caption{Bluetooth AFH Excluding Channels after Wi-Fi Congestion}
  \label{fig43}
 \end{figure}  

\begin{figure}
  \includegraphics[width=\linewidth]{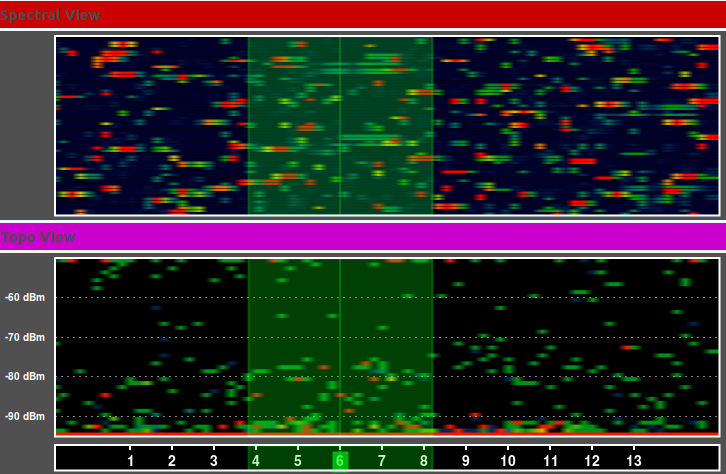}
  \caption{Bluetooth AFH re-uses channels previously excluded}
  \label{fig44}
 \end{figure}

To demonstrate the behaviour of Bluetooth’s AFH Mapping, the Wi-Fi traffic was stopped by switching off the Access Point, and disabling Wi-Fi on the smartphone being used for traffic generation. After around 10 seconds, the graphic shown in Figure \ref{fig43} was captured. At this point, Bluetooth audio has been playing continuously, and as the Wi-Fi traffic falls to zero, and scrolls off the top of the screen, the ‘gap’ where the Wi-Fi channel existed is shown to be empty of Bluetooth traffic as well. In the following seconds, as the Bluetooth devices test the channels previously obscured by Wi-Fi, these are re-used, and Bluetooth now occupies the full available spectrum (Figure \ref{fig44}).

This series of captures demonstrates AFH apparently working as expected, however, this is not as simple as it appears on initial inspection. Repeating the experiment and making slight changes to the order of events showed that, whilst Bluetooth’s AFH behaviour appears to behave as expected when Bluetooth starts communication in a heavily congested, saturated Wi-Fi environment, this is not the case when Bluetooth establishes a data connection in an environment with more typical Wi-Fi traffic – even if this traffic increases to cause congestion. In this circumstance, it appears that Bluetooth’s AFH behaviour does not work as expected, and instead, Wi-Fi and Bluetooth contest the available channels in a way that greatly reduces Wi-Fi throughput. During this sequence, the Wi-Fi throughput fell to as low as 1-2MBps, which was sustained until Bluetooth playback was eventually stopped.

\begin{figure}
  \includegraphics[width=\linewidth]{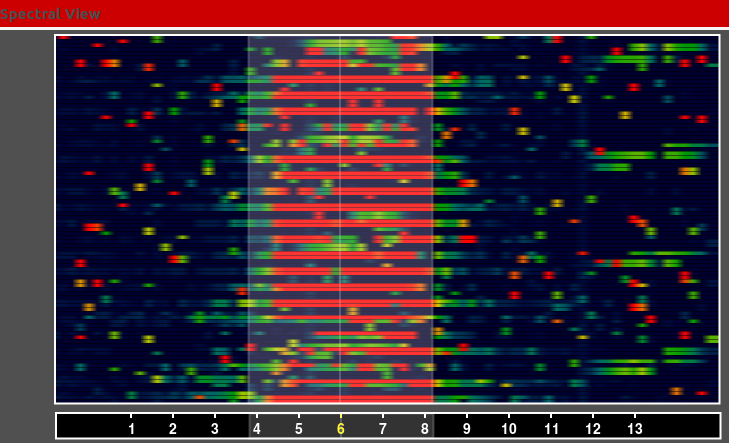}
  \caption{Bluetooth and Wi-Fi Interfering directly}
  \label{fig45}
 \end{figure}

This disputed use of the available channels by each technology is shown in Figure \ref{fig45} – the distinctive ‘banding’ pattern within the spectrum of Wi-Fi Channel 6 is the result of Bluetooth audio playback which started when Wi-Fi had not completely saturated the channel and Bluetooth has partial use of the spectrum. 

An AFHmap recovered using the Ubertooth tool during this test demonstrated that Bluetooth did not, in fact, consider the Wi-Fi channel’s space to be particularly ‘bad’ from a signal strength perspective. 

 \begin{figure}
  \includegraphics[width=\linewidth]{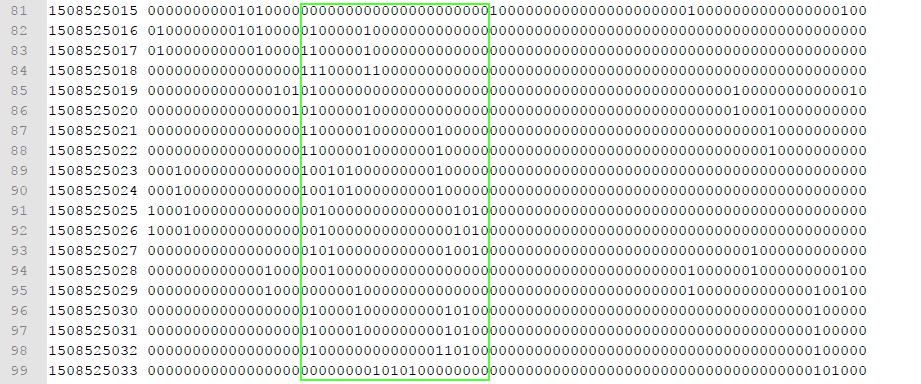}
  \caption{AFH Map captured during Run 4, OPO-i30-RFBusy}
  \label{fig46}
 \end{figure}    

It should be stressed that the RF events tracked in these two diagrams do not necessarily align – they are chosen to be broadly representative of what was observed during this experiment, however, the timestamps and resolution of each capture mechanism do not support a direct one to one comparison.

Some previous work has been carried out on co-existence between Bluetooth and Wi-Fi technologies. \cite{Neumeier2013} proposes a simulation model using raytracing techniques to examine Bluetooth and Wi-Fi in a shared space; however, this work focuses on varying the signal strength of interfering transmitters, and does not consider the potential role of each technology’s behaviour to spectrum use. The paper by \cite{Gummadi2007} used as a basis for the experimental method of this project also considers co-existence issues, and again, focuses on RF signal strength as the key factor. 

From the observed behaviour, it appears that whilst Wi-Fi devices seem to recognise the interference they encounter from Bluetooth traffic to be a sign of congestion, and back off accordingly, Bluetooth does not consider Wi-Fi traffic which emerges within its spectrum to necessarily designate a channel as ‘bad’.  It is possible that this conflicting behaviour has its origins in the design philosophy of each protocol – Wi-Fi has been an IEEE developed 802.x class protocol since its inception, and the concept of CSMA-CA (Carrier Sense, Multiple Access, Collision Avoidance) produces a behaviour which places a strong emphasis on being a “good citizen” when sending data. Bluetooth, on the other hand, uses FHSS technology in part for obfuscation, but primarily to be more resilient. AFH deepens this idea; AFH does not exclude other frequencies because it believes they are used by others as some form of sacrifice, but merely because using those channels may reduce its own resilience and the reliability of sending information. 

In fact, the Bluetooth Specification is explicit about this – AFH is used to reduce the impact of interference from other devices operating in the ISM Band. The impact of interference caused by Bluetooth traffic to other ISM users is not mentioned. As this competitive behaviour has not been described in the literature, this experiment was repeated, providing the same result each time; initiating Bluetooth communications in a space where Wi-Fi is already present, but not fully saturating the available bandwidth, appears to result in a contest for available bandwidth – with Bluetooth “winning” each time.

\section{Conclusions}
\subsection{Summary of the Work done}
This paper carried out a literature review to explore the current position of Security Research into the Bluetooth protocol. Bluetooth is a complex protocol – significantly more so than comparable 802.x family protocols – and this complexity requires the researcher to understand the interaction between the various functions of the PHY, MAC and LLC layers to a greater extent than that required to investigate Wi-Fi, for instance. The literature review attempts to strike a balance between exploring these functional blocks in sufficient detail to understand the work and contribution of key researchers, without becoming bogged down in the multiple options at each step. In the most recent paper included in the review \cite{Seri2017} describe the multiple interfaces and schemes support at each layer of the protocol provide such a complex attack surface that it is difficult to see how it could be reasonably secured.

Survey papers are reviewed, highlighting the multiple ways in which Bluetooth can be compromised. Despite this, there is a clear pattern in these exploits. Since Bluetooth’s introduction, potential weaknesses in the pairing, authentication and data transfer elements of the protocol have been identified by researchers. From the earliest of these, there has been an assumption made that the Frequency Hopping behaviour of Bluetooth is of limited value in protecting data \cite{Jacobsson2001}. Despite these repeated assumptions and assertions, a gradually building body of research has indicated that, far from being an easily circumventable technical trick, akin to ‘security through obscurity’, the Bluetooth channel hopping mechanism places significant obstacles in the path of a would-be eavesdropper. Those attacks which have been demonstrated rather than merely hypothesised require the attacker to participate in the Bluetooth network correctly at the RF and Baseband level, and rely on weaknesses in the higher levels of the protocol – weakness in key management, service authentication, etc.

A particular research direction has been developing since proposed by Dominic Spill \cite{Spill2007} – the ability to passively sniff Bluetooth traffic from the air, as can be easily done with Wi-Fi traffic.  Even with considerable advances, this has proven to be a difficult process, relying on hidden variables which have to be brute forced or recreated from recovered information. A great deal of work has been contributed by the team behind BlueEar \cite{albazrqaoe2016practical} at Michigan State University in particular a single researcher, Wahhab Albazrqaoe. These two key contributors have presented capture which can work under narrow, specific circumstances and have each proposed avenues of future research to build on their work.
This paper seeks to experimentally evaluate one of these proposed research avenues, by testing whether deliberately restricting the bandwidth used by Bluetooth can force its hopping behaviour to be simplified, reducing the number of channels to ultimately make brute force attacks simpler. An experimental approach was designed, repurposing work on Wi-Fi and Zigbee coexistence by \cite{Gummadi2007} to develop a repeatable way to experiment on Bluetooth in a congested RF environment. 
The tools developed by \cite{Spill2007} and improved by \cite{albazrqaoe2016practical} were used in a series of experiments based on the approach of \cite{Gummadi2007} to measure the time taken to acquire the clock index, one of the elements of ‘hidden’ information required to follow Bluetooth’s hopping behaviour, and the packets subsequently decoded.

\bibliographystyle{IEEEtran}
\bibliography{main}

\begin{thebibliography}{10}
\providecommand{\url}[1]{#1}
\csname url@samestyle\endcsname
\providecommand{\newblock}{\relax}
\providecommand{\bibinfo}[2]{#2}
\providecommand{\BIBentrySTDinterwordspacing}{\spaceskip=0pt\relax}
\providecommand{\BIBentryALTinterwordstretchfactor}{4}
\providecommand{\BIBentryALTinterwordspacing}{\spaceskip=\fontdimen2\font plus
\BIBentryALTinterwordstretchfactor\fontdimen3\font minus
  \fontdimen4\font\relax}
\providecommand{\BIBforeignlanguage}[2]{{%
\expandafter\ifx\csname l@#1\endcsname\relax
\typeout{** WARNING: IEEEtran.bst: No hyphenation pattern has been}%
\typeout{** loaded for the language `#1'. Using the pattern for}%
\typeout{** the default language instead.}%
\else
\language=\csname l@#1\endcsname
\fi
#2}}
\providecommand{\BIBdecl}{\relax}
\BIBdecl

\bibitem{Haartsen2000}
\BIBentryALTinterwordspacing
J.~C. Haartsen, ``{The Bluetooth Radio System},'' \emph{IEEE Personal
  Communications}, no. February, pp. 28--36, 2000. [Online]. Available:
  \url{http://citeseerx.ist.psu.edu/viewdoc/citations;jsessionid=EC2B8E030D8A95A43A4F0752A1100C74?doi=10.1.1.11.8115}
\BIBentrySTDinterwordspacing

\bibitem{Jacobsson2001}
\BIBentryALTinterwordspacing
M.~Jacobsson and S.~Wetzel, ``{Security Weaknesses in Bluetooth},'' in
  \emph{Topics in Cryptology - CT-RSA 2001}.\hskip 1em plus 0.5em minus
  0.4em\relax San Francisco: Springer, 2001, pp. 176--191. [Online]. Available:
  \url{https://link.springer.com/chapter/10.1007/3-540-45353-9{\_}14}
\BIBentrySTDinterwordspacing

\bibitem{Heffernan2001}
\BIBentryALTinterwordspacing
D.~Heffernan and G.~Leen, ``{Vehicles without wires},'' \emph{Computing {\&}
  Control Engineering Journal}, vol.~12, no.~5, pp. 205--211, oct 2001.
  [Online]. Available:
  \url{http://ieeexplore.ieee.org/lpdocs/epic03/wrapper.htm?arnumber=4062494
  http://digital-library.theiet.org/content/journals/10.1049/cce{\_}20010501}
\BIBentrySTDinterwordspacing

\bibitem{checkoway2011comprehensive}
S.~Checkoway, D.~McCoy, B.~Kantor, D.~Anderson, H.~Shacham, S.~Savage,
  K.~Koscher, A.~Czeskis, F.~Roesner, T.~Kohno \emph{et~al.}, ``Comprehensive
  experimental analyses of automotive attack surfaces.'' in \emph{USENIX
  Security Symposium}.\hskip 1em plus 0.5em minus 0.4em\relax San Francisco,
  2011, pp. 77--92.

\bibitem{cheah2017towards}
M.~Cheah, S.~A. Shaikh, O.~Haas, and A.~Ruddle, ``Towards a systematic security
  evaluation of the automotive bluetooth interface,'' \emph{Vehicular
  Communications}, vol.~9, pp. 8--18, 2017.

\bibitem{Shaked2005}
\BIBentryALTinterwordspacing
Y.~Shaked and A.~Wool, ``{Cracking the Bluetooth PIN},'' \emph{Proceedings of
  the 3rd international conference on Mobile systems, applications, and
  services - MobiSys '05}, pp. 39--50, 2005. [Online]. Available:
  \url{http://portal.acm.org/citation.cfm?doid=1067170.1067176}
\BIBentrySTDinterwordspacing

\bibitem{Spill2007}
\BIBentryALTinterwordspacing
D.~Spill and A.~Bittau, ``{BlueSniff: Eve meets Alice and Bluetooth},''
  \emph{WOOT '07 Proceedings of the first USENIX workshop on Offensive
  Technologies}, p.~10, 2007. [Online]. Available:
  \url{http://dl.acm.org/citation.cfm?id=1323276.1323281}
\BIBentrySTDinterwordspacing

\bibitem{Huang2014}
J.~Huang, W.~Albazrqaoe, and G.~Xing, ``{BlueID: A practical system for
  Bluetooth device identification},'' in \emph{Proceedings - IEEE INFOCOM},
  2014, pp. 2849--2857.

\bibitem{Dunning2010}
J.~P. Dunning, ``{Taming the blue beast: A survey of bluetooth based
  threats},'' \emph{IEEE Security and Privacy}, vol.~8, no.~2, pp. 20--27,
  2010.

\bibitem{Haines2010}
\BIBentryALTinterwordspacing
B.~Haines, ``{Bluetooth Attacks},'' in \emph{Seven Deadliest Wireless
  Technologies Attacks}.\hskip 1em plus 0.5em minus 0.4em\relax Elsevier, 2010,
  ch.~3, pp. 43--55. [Online]. Available:
  \url{http://linkinghub.elsevier.com/retrieve/pii/B9781597495417000035}
\BIBentrySTDinterwordspacing

\bibitem{Chokshi2010}
\BIBentryALTinterwordspacing
R.~Chokshi, ``{Yes! Wi-Fi and Bluetooth Can Coexist in Handheld Devices},''
  Marvell Semiconductor, Inc., Tech. Rep. March, 2010. [Online]. Available:
  \url{http://www.marvell.com/wireless/assets/Marvell-WiFi-Bluetooth-Coexistence.pdf}
\BIBentrySTDinterwordspacing

\bibitem{Pelzl2006}
\BIBentryALTinterwordspacing
J.~Pelzl and T.~Wollinger, ``{Security Aspects of Mobile Communication
  Systems},'' in \emph{Embedded Security in Cars}, K.~Lemke, C.~Paar, and
  M.~Wolf, Eds.\hskip 1em plus 0.5em minus 0.4em\relax Berlin/Heidelberg:
  Springer-Verlag, 2006, pp. 167--185. [Online]. Available:
  \url{http://link.springer.com/10.1007/3-540-28428-1}
\BIBentrySTDinterwordspacing

\bibitem{Ossmann2009}
M.~Ossmann and D.~Spill, ``{Building an All-Channel Bluetooth Monitor},'' in
  \emph{ShmooCon 09}, 2009, p. 102.

\bibitem{BluetoothSIG2007}
\BIBentryALTinterwordspacing
{Bluetooth SIG}, ``{Bluetooth 2.1},'' Bluetooth Special Interest Group, Tech.
  Rep. July, 2007. [Online]. Available:
  \url{https://www.bluetooth.com/specifications/adopted-specifications/legacy-specifications}
\BIBentrySTDinterwordspacing

\bibitem{Naggs2013}
\BIBentryALTinterwordspacing
T.~Naggs, ``{Ubertooth Mailing List},'' 2013. [Online]. Available:
  \url{https://sourceforge.net/p/ubertooth/mailman/message/31237673/}
\BIBentrySTDinterwordspacing

\bibitem{Chen2012}
L.~Chen, P.~Cooper, and Q.~Liu, ``{Security in Bluetooth Networks and
  Communications},'' in \emph{Wireless Network Security}, L.~Chen, J.~Ji, and
  Z.~Zhang, Eds.\hskip 1em plus 0.5em minus 0.4em\relax Beijing: Springer,
  2012, pp. 77--94.

\bibitem{albazrqaoe2016practical}
\BIBentryALTinterwordspacing
W.~Albazrqaoe, J.~Huang, and G.~Xing, ``{Practical Bluetooth Traffic Sniffing :
  Systems and Privacy Implications},'' in \emph{MobiSys 16 Proceedings of the
  14th Annual International Conference on Mobile Systems, Applications, and
  Services}.\hskip 1em plus 0.5em minus 0.4em\relax Singapore: ACM, 2016, pp.
  333--345. [Online]. Available:
  \url{http://dl.acm.org/citation.cfm?doid=2906388.2906403}
\BIBentrySTDinterwordspacing

\bibitem{Ryan2013}
\BIBentryALTinterwordspacing
M.~Ryan, ``{Bluetooth: With Low Energy Comes Low Security},'' \emph{Proceedings
  of the 7th USENIX Conference on Offensive Technologies}, p.~4, 2013.
  [Online]. Available: \url{http://dl.acm.org/citation.cfm?id=2534748.2534754}
\BIBentrySTDinterwordspacing

\bibitem{Albazrqaoe2011}
W.~Albazrqaoe, ``{a Study of Bluetooth Frequency Hopping Sequence: Modeling and
  a Practical Attack},'' Masters Thesis, Michigan State University, 2011.

\bibitem{Hodgdon2003}
\BIBentryALTinterwordspacing
C.~Hodgdon, ``{Adaptive Frequency Hopping for Reduced Interference between
  Bluetooth and Wireless LAN},'' 2003. [Online]. Available:
  \url{https://tinyurl.com/y86ztozk}
\BIBentrySTDinterwordspacing

\bibitem{Popovski2006}
P.~Popovski, H.~Yomo, and R.~Prasad, ``{Strategies for adaptive frequency
  hopping in the unlicensed bands},'' \emph{IEEE Wireless Communications},
  vol.~13, no.~6, pp. 60--67, 2006.

\bibitem{Tabassam2007}
\BIBentryALTinterwordspacing
A.~A. Tabassam, S.~Heiss, and M.~Hoing, ``{Bluetooth Device Discovery and Hop
  Synchronization by the Eavesdropper},'' in \emph{2007 International
  Conference on Emerging Technologies}.\hskip 1em plus 0.5em minus 0.4em\relax
  IEEE, nov 2007, pp. 1--5. [Online]. Available:
  \url{http://ieeexplore.ieee.org/document/4516305/}
\BIBentrySTDinterwordspacing

\bibitem{Scarfone2008}
\BIBentryALTinterwordspacing
K.~Scarfone, J.~Padgette, and L.~Chen, ``{Guide to Bluetooth security},''
  \emph{NIST Special Publication}, vol.~1, no. Guide to Bluetooth Security, p.
  121, 2008. [Online]. Available:
  \url{http://www.mcs.csueastbay.edu/{~}lertaul/BluetoothSECV1.pdf}
\BIBentrySTDinterwordspacing

\bibitem{BluetoothSIG2016}
\BIBentryALTinterwordspacing
{Bluetooth SIG}, ``{Bluetooth 5.0},'' Bluetooth Special Interest Group, Tech.
  Rep. December, 2016. [Online]. Available:
  \url{https://www.bluetooth.org/DocMan/handlers/DownloadDoc.ashx?doc{\_}id=421043}
\BIBentrySTDinterwordspacing

\bibitem{Spill2012}
\BIBentryALTinterwordspacing
D.~Spill, ``{Bluetooth Packet Sniffing Using Project Ubertooth},'' \emph{Ruxcon
  2012 Proceedings}, 2012. [Online]. Available:
  \url{http://2012.ruxcon.org.au/assets/rux/Spill-Ubertooth.pdf}
\BIBentrySTDinterwordspacing

\bibitem{CPUBenchmark2017}
\BIBentryALTinterwordspacing
{CPU Benchmark}, ``{CPU Comparison},'' 2017. [Online]. Available:
  \url{https://www.cpubenchmark.net/compare.php?cmp[]=3092{\&}cmp[]=1074}
\BIBentrySTDinterwordspacing

\bibitem{Rivertz2005}
\BIBentryALTinterwordspacing
H.~J. Rivertz, ``{Bluetooth Security},'' Norwegian Computing Centre, Oslo,
  Tech. Rep., 2005. [Online]. Available:
  \url{papers3://publication/uuid/0f81fe0b-3210-4140-80f2-9cb6bfe8ae44}
\BIBentrySTDinterwordspacing

\bibitem{Chernyshev2017}
\BIBentryALTinterwordspacing
M.~Chernyshev, C.~Valli, and M.~Johnstone, ``{Revisiting Urban War Nibbling:
  Mobile Passive Discovery of Classic Bluetooth Devices Using Ubertooth One},''
  \emph{IEEE Transactions on Information Forensics and Security}, vol.~12,
  no.~7, pp. 1625--1635, 2017. [Online]. Available:
  \url{http://ieeexplore.ieee.org/document/7872410/}
\BIBentrySTDinterwordspacing

\bibitem{Gummadi2007}
\BIBentryALTinterwordspacing
R.~Gummadi, D.~Wetherall, B.~Greenstein, and S.~Seshan, ``{Understanding and
  mitigating the impact of RF interference on 802.11 networks},'' \emph{ACM
  SIGCOMM Computer Communication Review}, vol.~37, no.~4, p. 385, 2007.
  [Online]. Available:
  \url{http://portal.acm.org/citation.cfm?doid=1282427.1282424}
\BIBentrySTDinterwordspacing

\bibitem{Muniz2013}
J.~Muniz and A.~Lakhani, \emph{{Web Penetration Testing with Kali
  Linux}}.\hskip 1em plus 0.5em minus 0.4em\relax Packt Publishing, 2013.

\bibitem{Neumeier2013}
R.~Neumeier and G.~Ostermayer, ``{Analyzing coexistence issues in wireless
  radio networks: Simulation of Bluetooth interfered by multiple WLANs},''
  \emph{Lecture Notes in Computer Science (including subseries Lecture Notes in
  Artificial Intelligence and Lecture Notes in Bioinformatics)}, vol. 8310
  LNCS, pp. 128--138, 2013.

\bibitem{Seri2017}
\BIBentryALTinterwordspacing
B.~Seri and G.~Vishnepolsky, ``{BlueBorne},'' Armis Inc., Tech. Rep., 2017.
  [Online]. Available: \url{http://go.armis.com/blueborne-technical-paper}
\BIBentrySTDinterwordspacing

\end{thebibliography}

\end{document}